\documentclass[twocolumn,showpacs,preprintnumbers,amsmath,amssymb]{revtex4-1}
\usepackage{graphicx}
\usepackage{dcolumn}
\usepackage{bm}
\usepackage{color}
\usepackage{amsmath, amssymb, gensymb}
\usepackage{subfigure}
\usepackage[svgnames*,table,dvipsnames]{xcolor} 
\usepackage{ulem}

\begin{document}
\preprint{APS/123-QED}

\title{Local and global avalanches in a 2D sheared granular medium}
\author{Jonathan Bar\'{e}s}\email{jb@jonathan-bares.eu}
\altaffiliation[Current address: ]{LMGC, UMR 5508 CNRS-University Montpellier, 34095 Montpellier, France}
\affiliation{Department of Physics \& Center for Nonlinear and Complex Systems, Duke University, Durham, North Carolina 27708, USA}
\author{Dengming Wang}
\affiliation{Key Laboratory of Mechanics on Western Disaster and Environment, Ministry of Education of China, Lanzhou University, 730000 Lanzhou, China}
\author{Dong Wang}
\affiliation{Department of Physics \& Center for Nonlinear and Complex Systems, Duke University, Durham, North Carolina 27708, USA}
\author{Thibault Bertrand}
\affiliation{Department of Mechanical Engineering and Materials Science, Yale University, New Haven, Connecticut 06520-8286, USA}
\author{Corey S. O'Hern}
\affiliation{Department of Mechanical Engineering and Materials Science, Yale University, New Haven, Connecticut 06520-8286, USA \\
Department of Physics, Yale University, New Haven, Connecticut 06520-8286, USA \\
Department of Applied Physics, Yale University, New Haven, Connecticut 06520-8286, USA}
\author{Robert P. Behringer}
\affiliation{Department of Physics \& Center for Nonlinear and Complex Systems, Duke University, Durham, North Carolina 27708, USA}

\begin{abstract}

We present the experimental and numerical studies of a 2D sheared amorphous material constituted of bidisperse photo-elastic disks. We analyze the statistics of avalanches during shear including the local and global fluctuations in energy and changes in particle positions and orientations. We find scale free distributions for these global and local avalanches denoted by power-laws whose cut-offs vary with inter-particle friction and packing fraction. Different exponents are found for these power-laws depending on the quantity from which variations are extracted. An asymmetry in time of the avalanche shapes is evidenced along with the fact that avalanches are mainly triggered from the shear bands. A simple relation independent from the intensity, is found between the number of local avalanches and the global avalanches they form. We also compare these experimental and numerical results for both local and global fluctuations to predictions from meanfield and depinning theories. 

\end{abstract}

\date{\today}
\pacs{81.05.Rm 45.70.Ht 91.30.Px 45.70.-n} 
\maketitle

\section{Introduction} \label{intro}

Yield-stress granular media flow if a sufficient shear stress is applied to them. Under certain conditions this flow can be spatially heterogeneous and intermittent in time. Although such behavior is widely observed in nature during avalanches, landslides and earthquakes, there are few quantitative experimental measurements of these intermittent dynamics at the local scale in frictional systems \citep{howell_prl1999,kabla2007_jfm,daniels2008_jgr,lebouil_prl2014}. Because of this, there is currently little understanding of the coupling between the evolution of meso-scale force chain networks and the macroscale mechanical response. This article aims to provide a first step in achieving this goal.

\begin{figure}[htb!]
\includegraphics[width=0.45\textwidth]{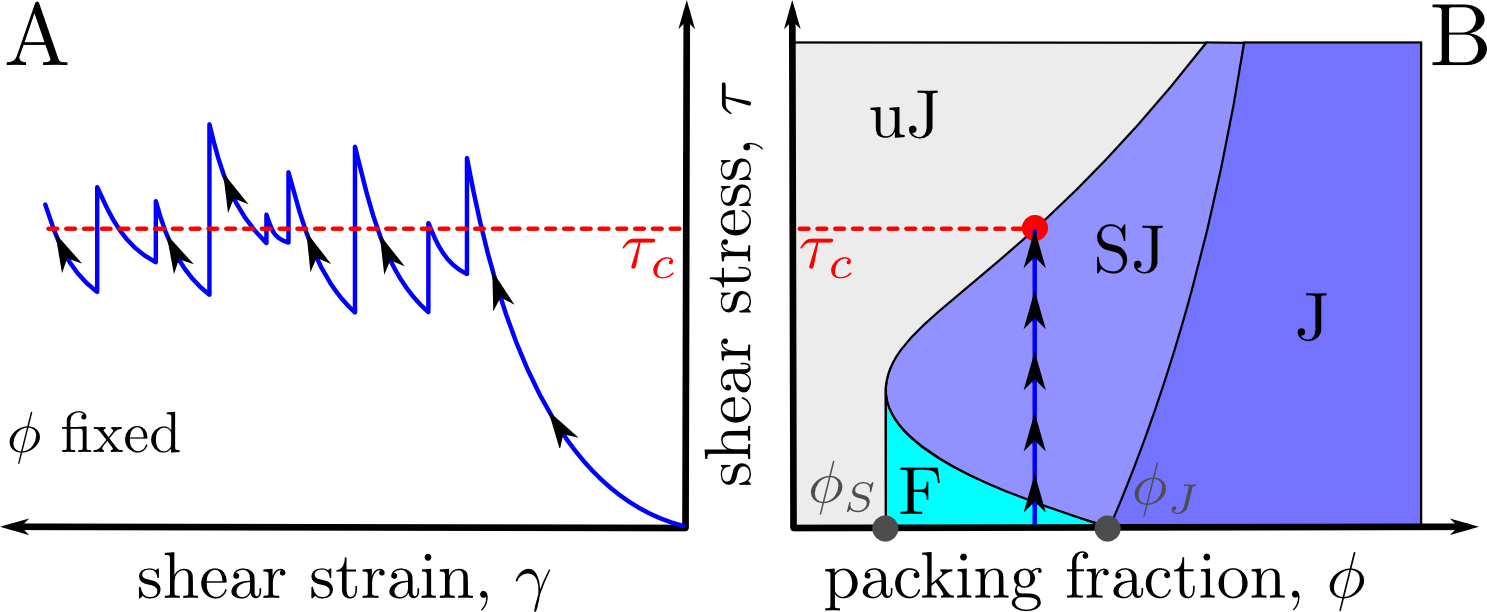}
\caption{(color online) A: Schematic view of the evolution of the shear stress $\tau$ in a sheared granular system at constant volume, for slowly loaded stiff grains ($x$-axis increases to the left). After a transient regime where the stress grows elastically, the shear stress fluctuates about the yield stress $\tau_c$, as observed in several experiments and simulations \citep{regev2015_nat,maloney2006_pre,dalton_pre2001,hartley_nat2003,kabla2007_jfm} B: Shear jamming phase diagram (adapted from \citep{bi_nat2011}) including jammed (J) \citep{liu1998_nat}, unjammed (uJ), fragile (F) \citep{cates1998_prl} and shear-jammed (SJ) regimes. A frictional granular system slowly sheared at constant packing fraction $\phi$ follows the vertical line with arrows until reaching the yield stress $\tau_c$, where it will fluctuate between jammed and unjammed states below and above $\tau_c$. The fluctuations above $\tau_c(\phi)$ give rise to intermittent dynamics (see A).}
\label{figIntroCrakling}
\end{figure}

Many experiments \citep{howell_prl1999,dalton_pre2001,hartley_nat2003,baldassarri_prl2006,petri_epjb2008,amon_prl2012,berhinger_jsm2014,lebouil_prl2014}, numerical simulations \citep{maloney_prl2004,aharonov_jgr2004,maloney_prl2009,salerno_prl2012,otsuki_pre2014} and models \citep{dahmen_nat2011,lin_pnas2014} involving systems that are subject to slow continuous loading, exhibit global intermittent dynamics characterized by a slow build-up and more rapid release of stress in the system. These systems include disordered molecular solids, metallic glasses, and granular materials, among others. A fundamental feature of dynamics in these systems is that the system remains near the yield stress curve, alternating between jammed and unjammed states \citep{Sethna_nat2001,Dickman_bjp2000}. The yield surface of granular materials is sensitive to the presence of inter-grain friction \citep{bi_nat2011}, and for a substantial range of the friction coefficient, $\nu$, the yield stress curve forms the upper boundary of the shear jamming phase diagram sketched in fig.\ref{figIntroCrakling}. When a frictional granular system at low pressure and shear stress, $\tau$, is sheared at constant volume, its global pressure, the Reynolds pressure \citep{reynolds_pm1885} increases. If the initial state was stress free, force chains appear and grow in the system \citep{Ren_prl2013} during this process; as shown in fig.\ref{figIntroCrakling}-B, as the shear stress increases, the system transitions from stress-free to fragile (F) and then from fragile to shear-jammed (SJ). When $\tau$, the shear stress in the system increases above the yield stress, $\tau_c (\phi)$ on the yield stress curve, the system unjams and $\tau$ falls to a jammed state with lower shear stress, $\tau < \tau_c$. In general, the shear stress at failure, and the shear stress to which the system returns following a failure are stochastic variables. These values of $\tau$ before and after the avalanche are both random variables, whose means determine the average yield stress. The yield stress curve, $\tau_c(\phi)$, represents an appropriate mean representation of the shear stress at failure for a given $\phi$. The rapid system evolution immediately following failure is an `avalanche', and a macroscopic measure of its strength can be the associated energy drop. However, the triggering of an avalanche occurs at a micro/mesoscopic scale. For systems that are close to force balance, due to the divergence of the length scale in the stress response of a jammed medium when approaching jamming from above \citep{wyart2005_arx}, a local failure can lead to a relatively long-ranged response in the force network, but relatively local changes in grain positions and orientations. 

In frictional granular materials, forces propagate nonuniformly along filamentary structures called force chains \citep{majmudar-nature-2005}. The stability of the force chains, hence the stability of jammed frictional granular materials, can be understood in terms of the number of contacts per particle, $Z$. In the vicinity of the jamming transition, it can be shown from generalized isostaticity \citep{shundyak-pre-2007} that the number of contacts per force-bearing particle is 3 ($Z\sim3$) in two dimensions for systems composed of highly frictional disks. Force chains are not typically straight; instead there is weak coupling between grains in a given force chains and neighboring chains. When a segment of force chain fails, {\it i.e.} when contacts between the grains constituting the segment fail, other neighboring force chains crossing the failed segment, or supported by the segment can fail as well. Thus, there exists an additional coupling between neighboring regions of the system. In the case of stress controlled dynamics, a failure in one part of the system will cause the force exerted by the boundaries elsewhere to increase, which leads to an additional coupling on a larger scale. As the present experiments and simulations are strain controlled, this additional boundary related coupling does not occur. However, the strong coupling of forces along chains, the weak coupling between chains, and the accompanying anisotropy in the stress and fabric tensors are key features of granular packings. 

Recently, Regev et al. proposed a mean field model to capture key features of the avalanche process \citep{regev2015_nat} for a range of systems exhibiting avalanches. In addition, studies of sheared amorphous materials commonly use molecular dynamic simulations and mean field theory; results of these numerical simulations in terms of avalanches of rearrangements have been contrasted to depinning models \citep{baret2002_prl,vandem2004_pre,alava2002_jpcm}. In more conventional amorphous solids, the microscopic plastic deformation, is thought to come from local rearrangements of particles involving Shear Transformation Zones (STZ) \citep{Langer2006_sm,manning2007_pre,bouchbinder2007_pre,lieou2012_pre,lieou2014_pre}. For these systems, described by linear isotropic elasticity, the result of a local failure is an Eshelby-like elastic field \citep{maloney2006_pre,falk1998_pre}. This mechanism, which describes the jamming-unjamming process as a kind of dynamic attractor, was suggested some time ago by \citep{cates1998_prl} in the conclusion of their paper invoking Self-Organized Criticality (SOC) \citep{Sethna_nat2001,Dickman_bjp2000,bak1988_pra}. In these models, the processes occur around a critical point \citep{biroli2007_nat}, and some observables undergo significant fluctuations, leading to the violence of the avalanche phenomenon and to power-law statistics.

We emphasize that the physical picture described in fig.\ref{figIntroCrakling} is specific to {\it sheared, frictional} granular systems. At the global scale, the material needs to be close to failure, and only shear provides such a state \citep{cates1998_prl} in the steady state regime. In the case of frictional granular materials, `failure' can occur either because the system is fragile or because it is driven across the yield stress surface. These are fundamentally different processes. Note that here, `fragile' refers to an instability under shear strain reversal and occurs in the region marked `F' in fig.\ref{figIntroCrakling}B. In this regime, if the direction of the shear strain is reversed at constant volume fraction $\phi$ from a direction that has established a weak network of force chains, all stresses, including the shear stress, drop substantially, possibly to zero, before a new network is established, and the stresses once again increase. This type of failure corresponds to a switch from a largest principal stress, $\sigma_1$, in one direction to a major principal stress $\sigma_2$, in a direction that is (nominally) orthogonal to the first. By contrast, failure at the yield surface occurs via a reduction in stresses that does not reverse the major principal stress direction. In both of these cases, the density/packing fraction typically remains fixed.  By contrast, in the compression case, $\phi$ increases; $\tau$ is not controlled, but compression tends to make the material more isotropic, and it is possible that $\tau = \sigma_1 - \sigma_2$ may decrease, fig.\ref{figIntroCrakling}-B. The loading rate, which is constant and slow enough to be in the quasistatic regime for the present studies, is an important physical parameter; however, the effect of finite loading rate is outside the scope of the present study and will be the object of future studies.

Since our grains are frictional, the jamming diagram (fig.\ref{figIntroCrakling}-B) presents a region of volume fraction below the isotropic jamming packing fraction for frictionless particles (volume fraction $\phi<\phi_J$), in which states ranging from stress-free to fragile, robustly shear jammed and flowing co-exist at the same time. The statistical behavior of the system differs depending on the driving along the shear stress direction in this phase diagram. Indeed, in stress-controlled protocols, the system can be loaded at constant stress above the yield stress ($i$) or with continuously increasing stress ($ii$); the system will not display a continuous avalanching regime in either of these cases. In case ($i$), the system becomes stuck after a transient regime ($\tau<\tau_c$) or never stops ($\tau_c<\tau$) and flows indefinitely. In case ($ii$), the system moves outside the jammed regime after a transient regime, and flows indefinitely. In strain-controlled experiments ($iii$), the statistical behaviour of the system reaches steady-state behavior after a transient regime; during this steady-state, the system oscillates around the yield stress curve. Although the dynamics of both regimes ($i$) and ($ii$) involve strong fluctuations, their statistics differ from those induced by ($iii$) where for ($iii$), the fluctuation dynamics remain unchanged as long as the strain is increased.

To understand, predict and potentially control the occurrence of avalanches, it is important to detect and track the physical mechanisms from the smallest scale, a particle size, where localized triggering occurs, to the system scale, where the effect of the avalanche is often detected. Hence, in this article we present experiments and numerical simulations where the full range of scales are studied. In both cases, we consider 2D granular materials consisting of bidisperse disks that are quasistatically sheared at constant volume fraction $\phi$. In particular, we track the energy and pressure stored in the system, as well as the particle-scale properties, including particle positions and rotations. We also present novel methods to measure the intensity and position of local and global avalanches. We then use these methods to determine statistical measures of avalanches and the inter-dependency of global and local events.

\section{Methods} \label{Setup}

\subsection{Experimental setup} \label{expe}

A typical experiment involves cyclically shearing a set of bidisperse 2D circular particles in a pure shear apparatus. Particles are photoelastic disks of thickness $6.35{\rm ~mm}$ and diameters $12.7$ and $15.9{\rm ~mm}$ (diameter ratio $d \approx 1.25$) made of Vishay PSM-4, as shown in fig.\ref{figParticle}. We use a bidisperse mixture to avoid crystallization, and the ratio between the number of small to large particles is kept constant at $3.3:1$ for all experiments. One set of particles is wrapped with Teflon$^{\textsc{\textregistered}}$ tape to reduce the friction coefficient between particles (see fig.\ref{figParticle}-D). The static friction coefficient is $\nu = 0.7$ and $0.2$ for unwrapped (bare) and wrapped particles, respectively. In order to track rotations, each particle is marked along its diameter with UV ink. 

\begin{figure}[htb!]
		\includegraphics[width=0.48\textwidth]{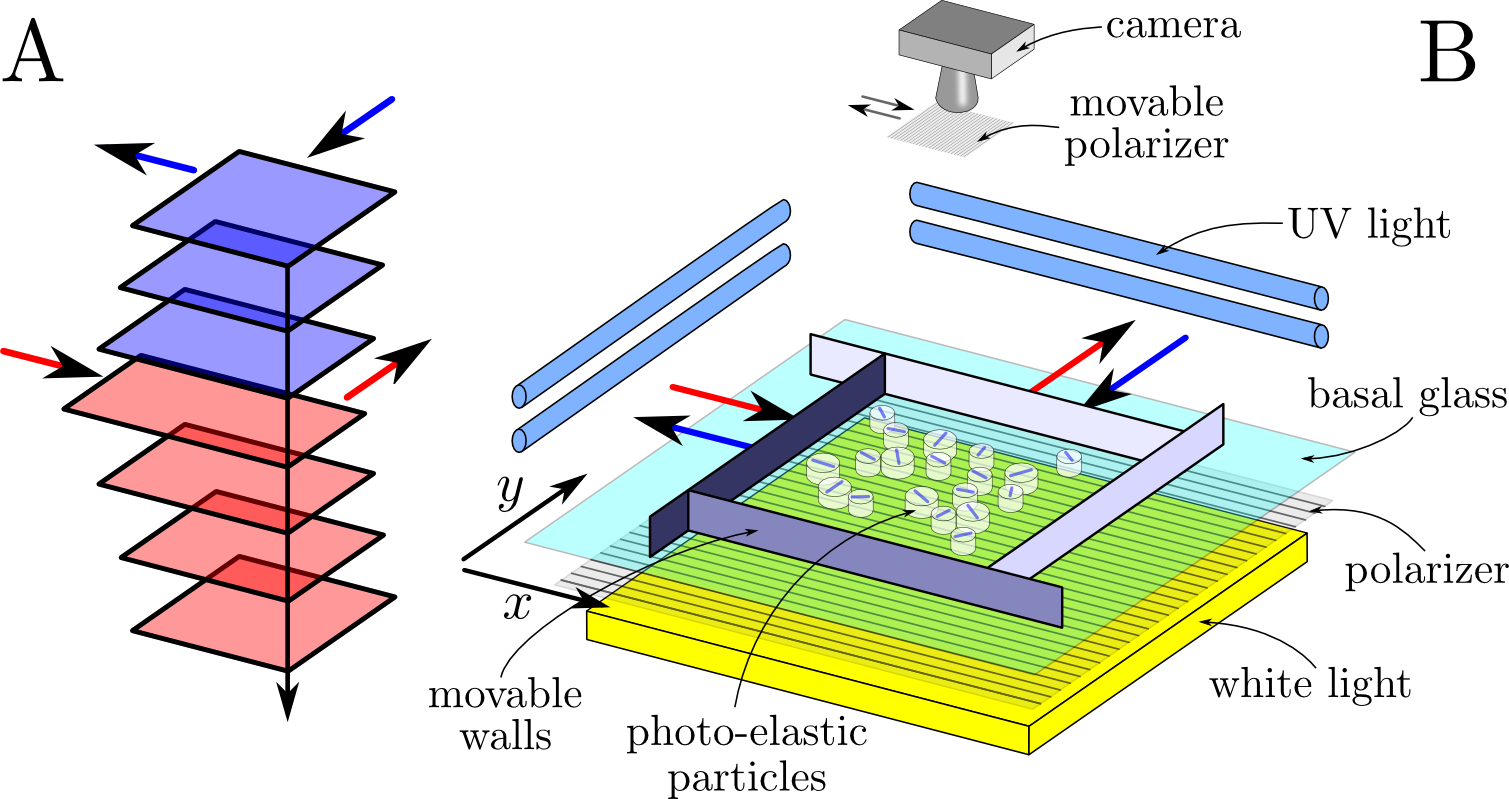}
\caption{(color online) A: From the top to bottom, evolution of the dimension of the pure shear cell during one cycle. An initial $44 \times 40{\rm~cm^2}$ rectangle is progressively shrunk in one direction and expanded in the other one to form a $55 \times 32{\rm~cm^2}$ rectangle (blue rectangles) and the motion is then reversed (red rectangles) to achieving a $20$\% shear amplitude. B: 3D schematic of the biaxial experimental cell. Moving walls shear, step by step, a set of bidisperse photo-elastic particles with a UV-ink bar on them. At each step the system is imaged with white light, between crossed polarizers and with UV light.}
\label{figExperiment}
\end{figure}

These particles rest on a transparent Plexiglas$^{\textsc{\textregistered}}$ plate slightly covered with talc to reduce the basal friction. The experiment is illuminated from below by a circularly polarized uniform white light and from above by a less intense UV light source. An $18$ megapixel SLR camera is placed $2{\rm~m}$ above the particles and can record pictures with and without the circular polarizer (see fig.\ref{figExperiment}-B). After each pure shear step, the system is imaged without the top polarizer (fig.\ref{figParticle}-A), with crossed polarizers (fig.\ref{figParticle}-B) and with the white light off and UV light on (fig.\ref{figParticle}-C). Experiments have been carried out for different packing fractions, for different shear amplitudes and for particles with different static friction coefficients as summarized in Table \ref{tabExpe}.

Pure shear strain (see geometry in fig.\ref{figExperiment}-A) is applied to the particle systems in small quasi-static steps, using the biaxial device shown in fig.\ref{figExperiment}-B. As showed in fig.\ref{figExperiment}-A, the boundaries of the cell compress the system in one direction and expand it in the other, keeping the area constant. Before each experiment, we prepare a stress-free packing of a given density by gently rearranging the particles. The initial boundary configuration is a $44 \times 40{\rm~cm^2}$ initial rectangle. During each experiment, this boundary spacing is shrunk by $1{\rm~mm}$ ($0.25\%$ strain) steps in the $y$-direction and expended in the $x$-direction to a $55 \times 32{\rm~cm^2}$ rectangle to reach a $20$\% shear amplitude (less for some high density experiments), while keeping the overall area constant.  The directions of compression and dilation are then reversed back to the initial boundary configuration. For each experiment, such a back and forth cycle is repeated $50$ times. During each step, (i) the boundary walls move for $2$s, (ii) the system is allowed to relax and (iii) the imaging process is carried out which last for $\sim10$s in total. The loading is slow enough to be considered as quasistatic.

\begin{table}[htb!]
	\begin{center}
		\begin{tabular}{|c|c|c|c|c|c|}
		\hline 
		experiment & $I$ & $II$ & $III$ & $IV$ & $V$ \\ 		
		\hline 
		particles & bare & bare & bare & bare & wrapped \\ 
		\hline 
		\begin{tabular}{c} packing \\ fraction \\ $\phi$ \end{tabular} & $0.785$ & $0.790$ & $0.799$ & $0.805$ & $0.808$ \\ 
		\hline 
		\begin{tabular}{c} shear \\ amplitude \end{tabular} & $20$\% & $17.5$\% & $10$\% & $7.5$\% & $20$\% \\ 
		\hline
		\end{tabular} 
	\end{center}
\caption{Input parameters of the experiments. Particles can be bare or wrapped with Teflon$^{\textsc{\textregistered}}$ tape to change their static friction coefficient. The total number of particles is changed to vary the system density but the number ratio between small and large particles stays the same. The maximum shear amplitude is chosen so that the pressure inside the system is low enough for it not to buckle.}
\label{tabExpe}
\end{table}

\subsection{Image post-processing} \label{imgPostP} 

\begin{figure}[htb!]
		\includegraphics[width=0.38\textwidth]{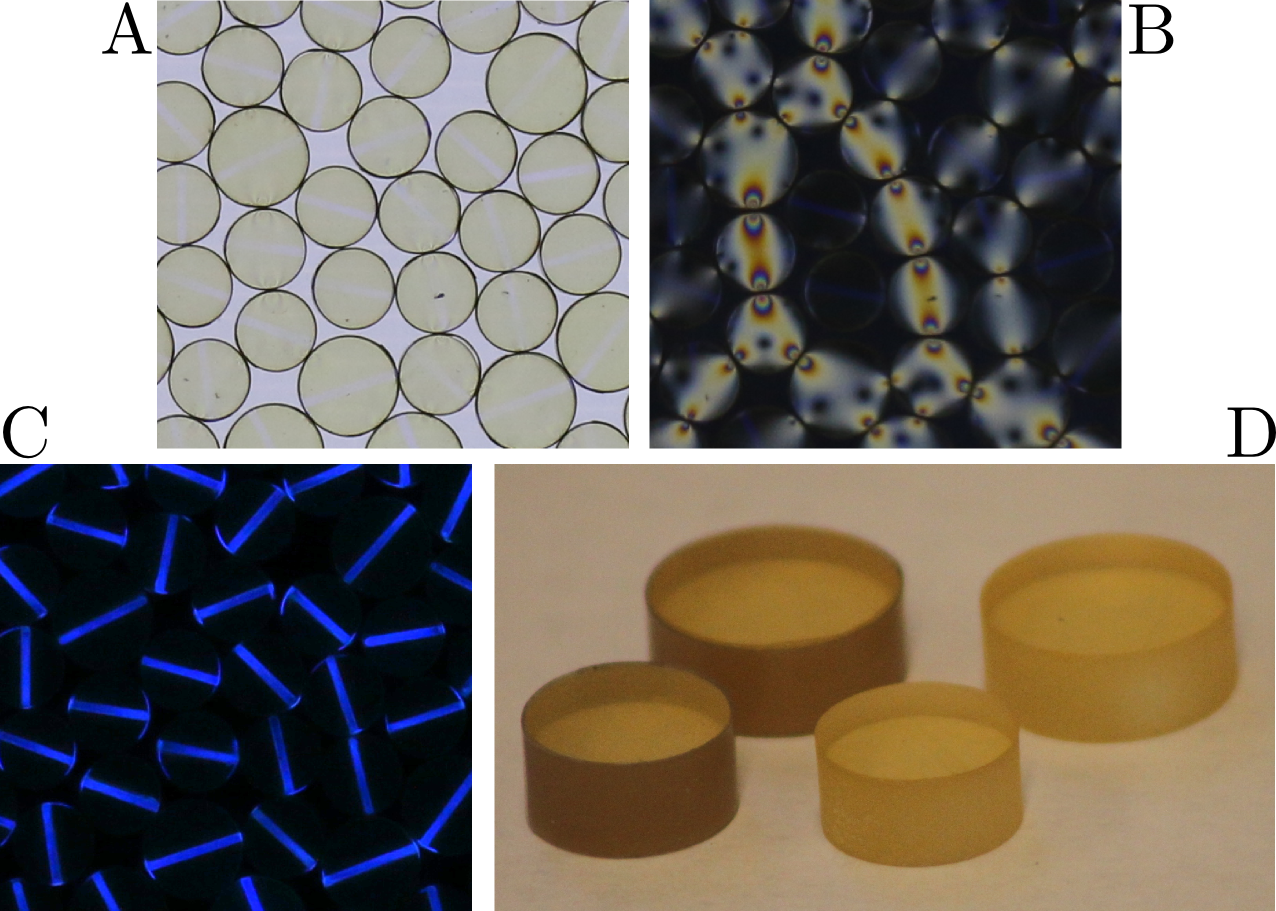}
\caption{(color online) Top view of the granular system in transmitted white light (A), between crossed polarisers (B) and in UV light (C). D: From left to right small and large Teflon$^{\textsc{\textregistered}}$ wrapped particles and small and large bare particles.}
\label{figParticle}
\end{figure}

As shown in fig.\ref{figParticle}-A, without the top crossed polarizer, the color of the particles differs from the background (yellowish). Using this property, the unpolarized pictures are converted to a binary representation (black for the particles and white for the background) with an adaptive threshold algorithm and convolved with a disk of the size of the particles for both particle diameters. The maximum of the convolutions for each diameter gives the particle position. From the cross-polarized pictures, we measure the pressure of each particle, using an empirical approach introduced in \citep{howell_prl1999}. If a quasi-2D photo-elastic object is observed between crossed polarizers, then for a given wavelength, the fraction of the light going through a portion of the material subjected to a local shear stress $\tau=\sigma_1-\sigma_2$ has intensity:

\begin{equation} \label{eqLightIntensity}
I \sim \sin^2\left(\dfrac{\pi CT}{\lambda}(\sigma_1-\sigma_2)\right),
\end{equation}

\noindent 
where $\sigma_1$ and $\sigma_2$ are principal stresses, $C$ is the material stress optic coefficient, $T$ is the object thickness (here $6.35$mm) and $\lambda$ is the wavelength of the light ($\sim 510$nm for the green filter we use). To obtain information on grain pressure, we use the fact that the contact forces acting on a grain create stresses inside the grain, which changes the phase variable, $\pi CT (\sigma_1-\sigma_2)/\lambda$, inside the sine function of eq.-\ref{eqLightIntensity}. Where the phase variable is an integer multiple of $\pi$, the corresponding transmitted image region is dark, and where the phase is an odd multiple of $\pi/2$ it is bright. In a photo-elastic image of a grain, increasing applied contact forces increases the stresses (both pressure and shear stress) within the grain, and leads to an increasing density of light and dark fringes. Since the pressure is a reflection of the mean normal forces on a particle, hence the internal stress, it is straight forward to calibrate a measure of the fringe density against the pressure, $P$ \citep{howell_prl1999,geng_prl2001,Ren_prl2013,berhinger_jsm2014}. To quantify the fringe density we measure the squared gradient of the photoelastic image intensity, $G^2$, integrated over a particle. This quantity provides an empirical connection to the local pressure acting on the grain. Since the material is purely elastic, the energy $e$ stored in the particle is proportional to $P^2$ or $G^4$.

The orientation of each particle is also measured using UV light imaging (fig.\ref{figParticle}-C) after each strain step. A Hough transform is performed locally on the binarized UV image of each grain to detect the fluorescent bar. An angle $\theta$ between $0$ and $\pi$ is then attributed to each grain for each step and a variation is deduced: 

\begin{equation}
\Delta \theta_i = |\theta_i(\gamma) - \theta_i(\gamma+\Delta \gamma)|
\end{equation}

\subsection{Numerical simulations} \label{NumSimul}

We also carry out corresponding Discrete Element Modeling (DEM) simulations. We employ a contact force model first developed by Cundal and Strack to describe the mechanical behavior of disks \citep{cundall1979_geo}, and more recently revised by Silbert et al. \citep{silbert2001_pre}. Here, we perform these numerical simulations using the DEM code LIGGGHTS initially developed by \citep{kloss2012_pcfd}. In the Hertz-Mindlin contact model, the grain-grain and grain-wall interactions are modeled using a spring-dashpot description. The contact forces, both normal and tangential, are represented by a purely repulsive Hertzian spring model with velocity-dependent damping. To maintain close contact with the experiments, we use a bidisperse mixture of disks with diameter $13$ and $16{\rm~mm}$ (diameter ratio $d=1.23$). We vary the packing fraction by adjusting the number of disks in the simulation cell, and the particles are loaded in the same geometry as the one described in fig.\ref{figExperiment}. Normal and tangential forces acting on a given particle $i$ from a particle $j$ are given by:

\begin{align}
&\mathbf{F}^{n}_{ij} = k_n\delta\mathbf{n}_{ij}-\gamma_n\mathbf{v}^n_{ij} \\
&\mathbf{F}^{t}_{ij} = k_t\delta\mathbf{t}_{ij}-\gamma_t\mathbf{v}^t_{ij}
\end{align}

\noindent where $k_n$ and $k_t$ are the stiffnesses for the normal and tangential springs, $\gamma_n$ and $\gamma_t$ are the viscoelastic damping constants for normal and tangential contacts, $\delta \mathbf{n}_{ij}$ and $\delta \mathbf{t}_{ij}$ are the normal and tangential displacement vectors between particles $i$ and $j$, and $\mathbf{v}^n_{ij}$ and $\mathbf{v}^t_{ij}$ are the relative normal and tangential velocities between particles $i$ and $j$.  To mimic the basal friction present in the experiments, the particles are also subjected to viscous fluid damping in the plane of motion.

The grain properties are set to closely match the experimental values. We use a Young's modulus $E = 4{\rm~MPa}$, a Poisson's ratio $\mu=0.49$, a density $\rho = 2500 \mathrm{kg.m}^{-3}$ and a coefficient of restitution $c_r=0.3$, close to the experimental material. We vary the friction coefficient common to grain-grain and grain-wall contact interactions $\nu$, as well as the packing fraction $\phi$ by changing the number of particles for a given load cell geometry. All of these conditions are summarized in Table \ref{tabNum}. In order to implement Coulomb static friction, we truncate the tangential displacement to fulfill the Coulomb sliding condition at each contact:

\begin{equation}
F^t_{ij} \le \nu F^n_{ij}
\end{equation}

\noindent Appendix \ref{app:coefficients} details the relationship between materials properties and the elastic and viscoelastic damping constants used in the simulations.

We obtain stress-free initial configurations at each volume fraction by isotropically growing particles randomly seeded in the load cell at very low density. After each growth step, we minimize the total potential energy using molecular dynamics with viscous damping. We then load the cell at a constant shear rate $\dot{\gamma}=10^{-5}$ which is in the quasistatic limit.

\begin{table}[htb!]
	\begin{footnotesize}		
			\begin{tabular}{|c|c|c|}
				\hline 
				simulation & \begin{tabular}{c} static friction coefficient \\ $\nu$ \end{tabular} & \begin{tabular}{c} packing fraction \\ $\phi$ \end{tabular} \\ 
				\hline 
				1 & 0.7 & 0.780 \\ 
				\hline 
				2 & 0.7 & 0.784 \\ 
				\hline 
				3 & 0.7 & 0.788 \\ 
				\hline 
				4 & 0.7 & 0.792 \\ 
				\hline 
				5 & 0.7 & 0.796 \\ 
				\hline 
				6 & 0.7 & 0.800 \\ 
				\hline 
				7 & 0.7 & 0.805 \\ 
				\hline 
				8 & 0 & 0.788 \\ 
				\hline 
				9 & 0.1 & 0.788 \\ 
				\hline 
				10 & 0.2 & 0.788 \\ 
				\hline 
				11 & 0.3 & 0.788 \\ 
				\hline 
				12 & 0.4 & 0.788 \\ 
				\hline 
				13 & 0.5 & 0.788 \\ 
				\hline 
				14 & 0.6 & 0.788 \\ 
				\hline 
				15 & 0.8 & 0.788 \\ 
				\hline 
				16 & 0.9 & 0.788 \\ 
				\hline 
				17 & 1 & 0.788 \\ 	
				\hline 
			\end{tabular} 			
	\end{footnotesize}
\caption{Input parameters of the numerical simulations. The static friction coefficient between particles is varied from perfectly slippery ($\nu=0$) to highly frictional ($\nu=1$). For each static friction coefficient that was near an experimental values, the total number of particles was changed to vary the system density but the ratio between small and large particles stayed the same.}
\label{tabNum}
\end{table}

\section{Avalanche detection} \label{AvlDetect}

\subsection{Measurement of global avalanches} \label{extrGlobal}

\begin{figure}[htb!]
		\includegraphics[width=0.5\textwidth]{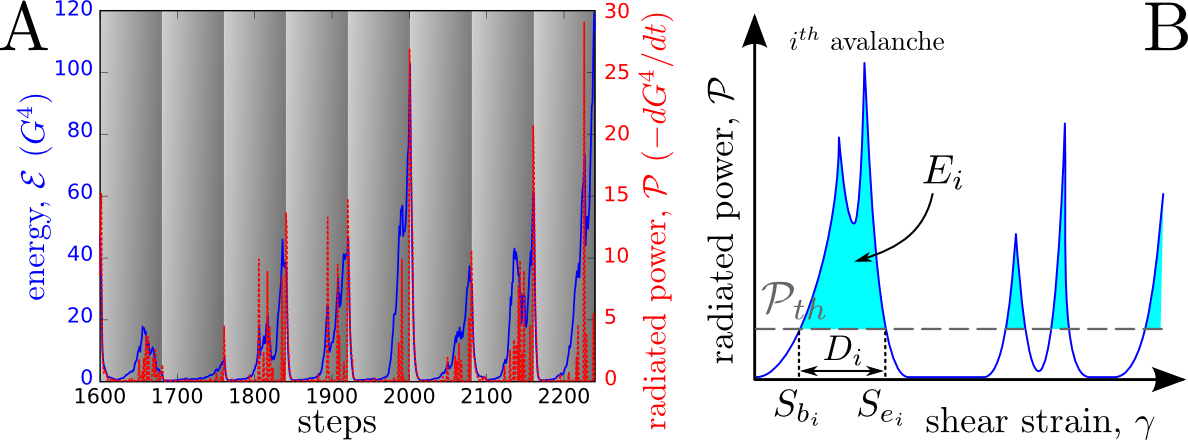}
\caption{(color online) A: The plain blue line is the global energy $\mathcal{E}$ stored in the granular media measured experimentally as the sum of $G^4$ over the particles. The red dashed line is the power $\mathcal{P}$ dissipated in the system computed as minus the negative part of the derivative of the energy with respect to strain. Both quantities are plotted as a function of the shear step. Each gray shaded box marks shearing in one direction (shearing is reversed periodically). B: Cartoon of the threshold power (energy released per strain unit) $\mathcal{P}$ signal to extract global avalanches. For each $i$ avalanche we measure the beginning strain ($S_{b_i}$), ending strain ($S_{e_i}$), duration ($D_i$) and accumulated dissipated energy ($E_i$).}
\label{figPower}
\end{figure}

From the sum of the particle energies $e_j$, we compute the evolution of the global energy $E=\sum\limits_{j} e_j$ stored in the granular system (see fig.\ref{figPower}-A). As expected from fig.\ref{figIntroCrakling}-A, we observe fluctuations of the signal due to two different features: very large fluctuations following reversal of the shear direction, but also large spontaneous fluctuations ({\it i.e.} avalanches) due to rearrangement of the grains inside the system.

We detect avalanches, measure their size and location, and analyze their energy drops using conventional pinning-depinning model approaches \citep{bares_ftp2014}. As shown in fig.\ref{figPower}-A, we compute the negative derivative of the energy with respect to strain ($\mathcal{P}=-d\mathcal{E}/d\gamma$), and consider only the positive part -- corresponding to the released energy. Then, as in fig.\ref{figPower}-B, we choose a threshold $\mathcal{P}_{th}$ and consider all peaks above this threshold as avalanches. For each peak $i$, the strain $S_{b_i}$ when the signal crosses the threshold going up is considered as the beginning of the avalanche, while the end corresponds to the signal crossing the threshold going down, $S_{e_i}$. The strain difference between those two events is the `duration' $D_i=S_{e_i}-S_{b_i}$ (in units of strain rather than time). The `size'  (in energy) of the avalanche or energy drop $E_i$ is given by the area under the peak and above the threshold value as presented in fig.\ref{figPower}-B. We do not consider events that are caused by the periodic reversal of the shear direction, since these are not spontaneous events. We note that avalanches are detected using the power (energy released by strain unit) instead of the pressure derivative. Characterizing avalanches in terms of an extensive quantity (energy, area...) is better than measuring it in terms an intensive one (pressure, force...) because the former does not intrinsically depend on any other quantity in the system such as area of contact, material, stiffness...

\subsection{Measurement of the local avalanches} \label{extrLocal}

\begin{figure}[htb!]
		\includegraphics[width=0.42\textwidth]{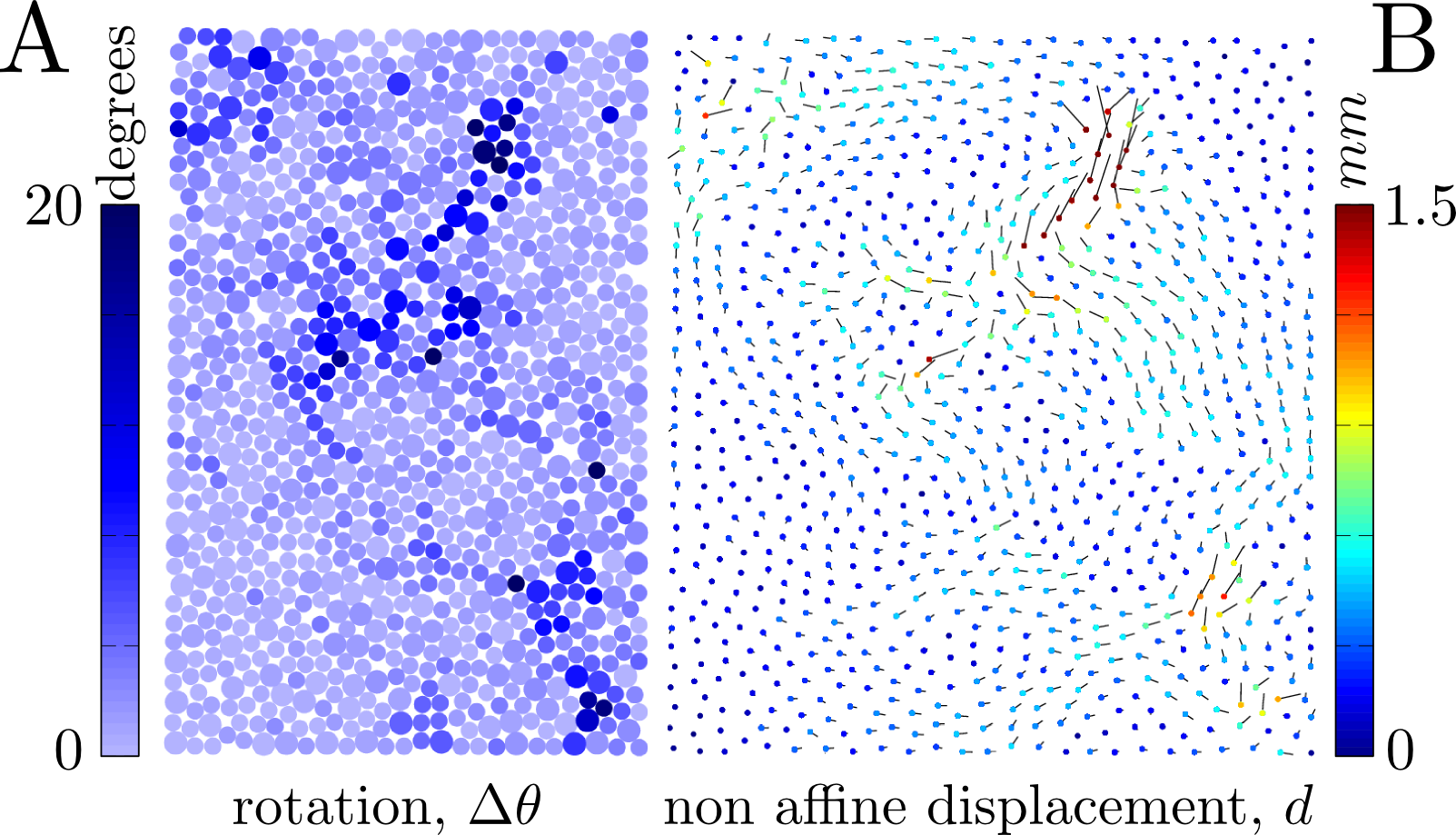}
\caption{(color online) A: Absolute value of the particle rotation $\Delta \theta$ measured from step $j_0$ to $j_0+1$ in experiment $I$. The clusters of large rotation corresponds to local grain rearrangements. B: Non-affine displacement of grains between step $j_0$ and $j_0+1$ (motion due to the boundary is removed). Particles positions after removing the affine displacements, interpolated from the boundary motion. The direction of non-affine motion is indicted by arrows and the magnitude of non-affine motion is indicated by the color scale. }
\label{figLocalAngleDisp}
\end{figure}

\begin{figure}[htb!]
		\includegraphics[width=0.49\textwidth]{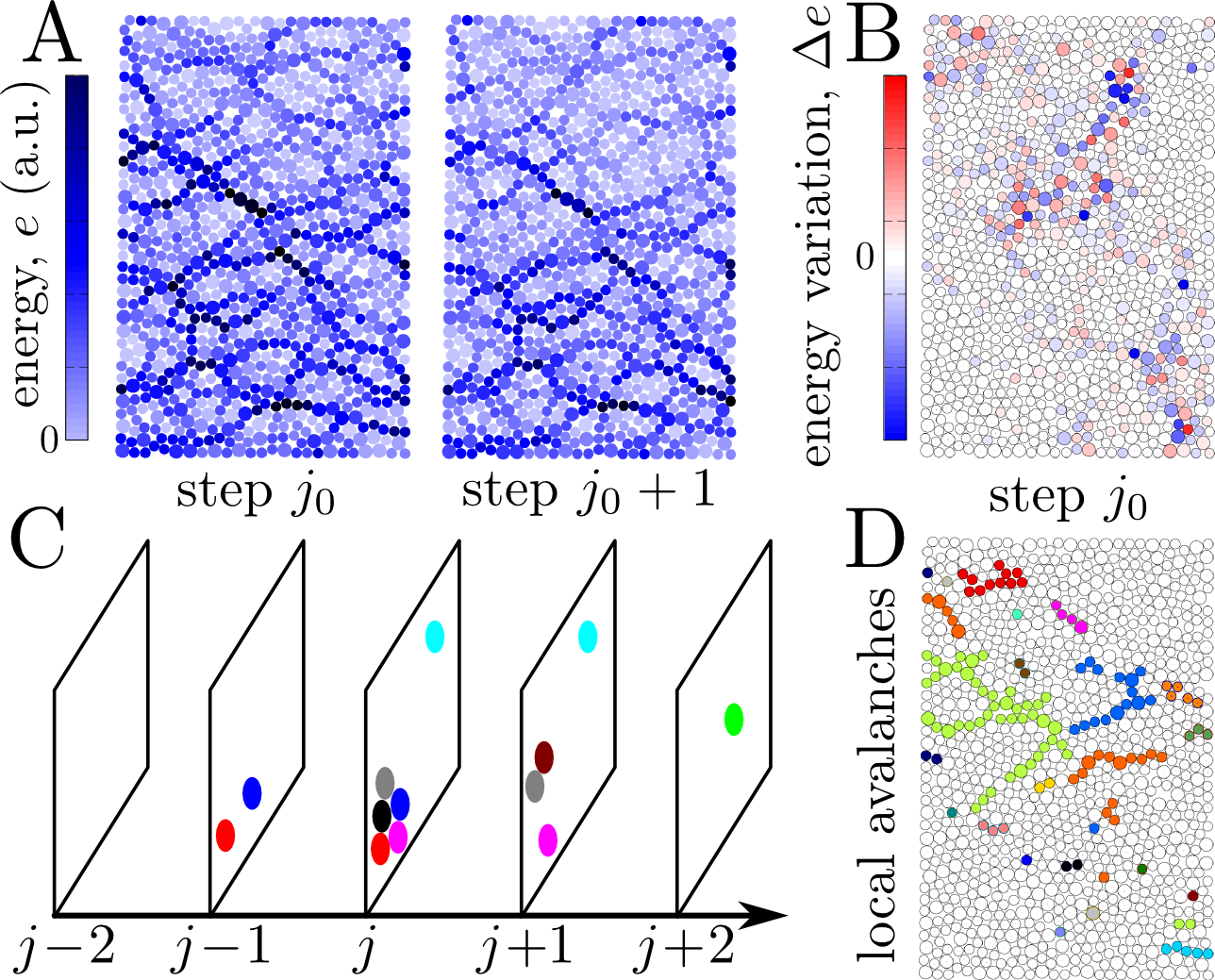}
\caption{(color online) A: Energy ($G^4$, arbitrary units) stored in each particle before (step $j_0$) and after (step $j_0+1$) an avalanche for experiment $I$. Step $j_0$ corresponds with fig.\ref{figLocalAngleDisp}. We notice that some force chains break. B: Variation of the energy from step $j_0+1$ to $j_0$. Blue particles are unloaded (lose energy) whereas red are loaded (gain energy). C: Schematic of the evolution of particle undergoing a strong energy drop after a shear step. Colors stand for the same particles from step to step. D: Grains involved in an avalanche at shear strain step $j_0$. Same colors stand for same avalanches. We see that local avalanches detected with energy follow the force chain structure. (see text for details)}
\label{figLocalExtraction}
\end{figure}

Avalanches at the global scale are created by rearrangements of the grains and the granular force network triggered by structural evolution at the local scale. Here, we define and detect those rearrangements which we call `local avalanches'. As in fig.\ref{figLocalAngleDisp}-A,B and fig.\ref{figLocalExtraction}-B, when the global energy varies strongly, particles in localized clusters undergo strong rotations, displacements and/or energy/pressure variations respectively. These dynamical heterogeneities are reminiscent of the ones already observed in \citep{dauchot_prl2005} and \citep{maloney_prl2004} respectively. Here, we isolate these sharp variations both in space and strain, and quantify their `size' and `duration'.

Local avalanches can be extracted from the pressure/energy fields (fig.\ref{figLocalExtraction}-A and B), the rotation field (fig.\ref{figLocalAngleDisp}-A) and the displacement field (fig.\ref{figLocalAngleDisp}-B). In this last case, we only consider the non-affine displacement, which is measured as the displacement corrected by the affine displacement imposed by the pure shear motion of the boundaries. For each of those quantities, at each shear step, we consider the particles with a value (energy drop for instance) higher than a certain threshold. In the schematic of fig.\ref{figLocalExtraction}-C for example, $6$ particles are involved in a local avalanche at step $j$, whereas no particles are involved in an avalanche at step $j-2$ (no significant rearrangement). Then, for each step we identify these particles as part of a single cluster if they are all in mutual contact. In fig.\ref{figLocalExtraction}-C at step $j$, we identify $2$ clusters. One is formed by the red, pink, blue, black, and grey particles and the other includes only the cyan particle. Finally, for the previous and successive steps, $j-1$ and $j+1$, we look for clusters such that at least one particle belongs to a cluster detected at step $j$. If we identify one, we connect these clusters in time to form an avalanche with duration $D_i$ as in fig.\ref{figLocalExtraction}-D at step $j_0$ for the energy drop. In fig.\ref{figLocalExtraction}-C, an avalanche formed from the red, pink, blue, black, grey and brown particles is created at step $j-1$ and ends at step $j+1$.

As for global avalanches, for each local avalanche $i$ detected with this method, we define:
\begin{itemize}
	\item The strain at the beginning $S_{b_i}$: the step where the first particle involved in the avalanche crosses the threshold.
	\item The strain at the end $S_{e_i}$: for the next step after that, the last particle involved in the avalanche crosses the threshold for the last time.
	\item The `duration' $D_i$: the difference between the ending and beginning strain ($D_i=S_{e_i}-S_{b_i}$)
	\item The position $(X_i,Y_i)$: the center of mass of all the particles involved in the avalanche, weighted by the number of times they are above the threshold.
	\item The intensity: the number of particles involved in the avalanche $N_i$ ($6$ in the example of fig.\ref{figLocalExtraction}-C) or as the total energy/pressure drop/change, rotation or displacement $E_i$, summed over all the steps and all the particles of the avalanche.
\end{itemize}
\noindent As explained for global avalanches, the best way to define an avalanche is to use the energy drop, rotation or displacement because they are extensive quantities so the measurement does not depend on any other quantity in the system. In the rest of this paper, we will mainly use the energy drop as the fluctuating quantity that defines the avalanche.

\section{Results} \label{result}

\subsection{Global avalanches} \label{resGlobal} 

\begin{figure}[htb!]
		\includegraphics[width=0.35\textwidth]{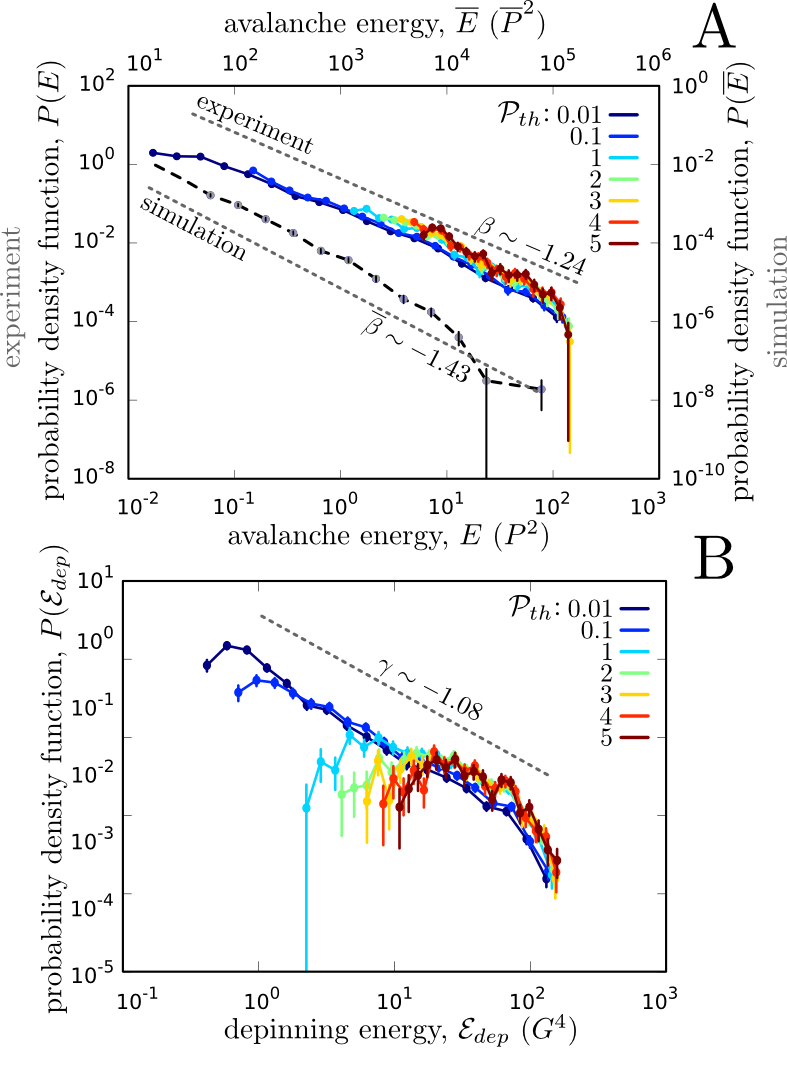}
\caption{(color online) A: Probability density function of the energy of global avalanches $P(E)$ for different threshold values $\mathcal{P}_{th}$ (color from blue to brown) for experiment $I$ and with a low threshold value ($\overline{\mathcal{P}_{th}}=10$) for simulation $2$ (black). For experimental data, the statistics follow a power-law over $3$ decades with exponent $\beta=-1.24 \pm 0.11$. For simulations, a power-law is also observed over $2.5$ decades with $\overline{\beta}=-1.43 \pm 0.14$. B: Probability density function of the energy stored in the system ($\mathcal{E}$) when an avalanche is triggered ($\mathcal{E}_{dep}$, depinning energy) for different threshold values for experiment $I$.}
\label{figPdfEloc}
\end{figure}

The probability density function (PDF) of the avalanche size is a standard tool used for analysing the dynamics of a system displaying crackling behaviour. For a typical experiment and simulation, we plot in fig.\ref{figPdfEloc}-A the energy, $E$, extracted from the global power signal, $\mathcal{P}$, for both experiment $I$ (see Table~\ref{tabExpe}) and numerical simulation $2$ (see Table~\ref{tabNum}). We note that in the rest of this paper, we will use Arabic numbering for simulations and Roman numbering for experiments. We find that these PDFs follow a power-law with exponent $\beta$:

\begin{equation} \label{eqPdfE}
P(E) \sim E^{\beta}
\end{equation}

\noindent The value of the exponent measured in the experimental case is $\beta=-1.24 \pm 0.11$ while the simulations yield $\overline{\beta}=-1.43 \pm 0.14$. We note that in the rest of this paper, quantities with a bar are from simulations and ones without a bar are from experiments. Although these exponents are somewhat different, they agree within the $95\%$ error bars. In the experimental case, we also explored the effect of the avalanche detection threshold $\mathcal{P}_{th}$ on the statistics; no effect on the exponent or the upper cut-off of the PDF was observed. This threshold only affects the lower cut-off because small avalanches are below the threshold and are not detected for high $\mathcal{P}_{th}$ values. Hence, to compare the statistical behavior of different systems, it is important to keep this parameter constant and low enough to avoid missing small events.

When an avalanche is triggered, not only the energy $E$ released by the system is relevant to characterize the avalanche behavior, but also $\mathcal{E}_{dep}$, the total energy stored in the system. Fig.\ref{figPdfEloc}-B shows the PDF of $\mathcal{E}_{dep}$ for experiment $I$ at strain $S_{b}$ when an avalanche is triggered. This energy is sometimes called the depinning energy, and in the framework of the pinning-depinning theory, several models predict Gaussian statistics \citep{bolech2004_prl,roux2008_ijf,talamali2011_pre,patinet2013_prl}. In our case, unlike these models, we observe a power-law distribution, and the measured experimental exponent is close to $1$: $\gamma=-1.08 \pm 0.1$. As the detection threshold, $\mathcal{P}_{th}$, increases at fixed upper cut-off, the distribution tends to a Gaussian distribution. This suggests that only large avalanches statistics can be described within the pinning-depinning framework. Moreover, the fact that only the lower cut-off is changed by the threshold value means that unlike other different crackling systems \citep{bares_prl2015}, by increasing $\mathcal{P}_{th}$ large avalanches do not break up into smaller ones and just more small avalanches will be excluded.

\begin{figure}[htb!]
		\includegraphics[width=0.48\textwidth]{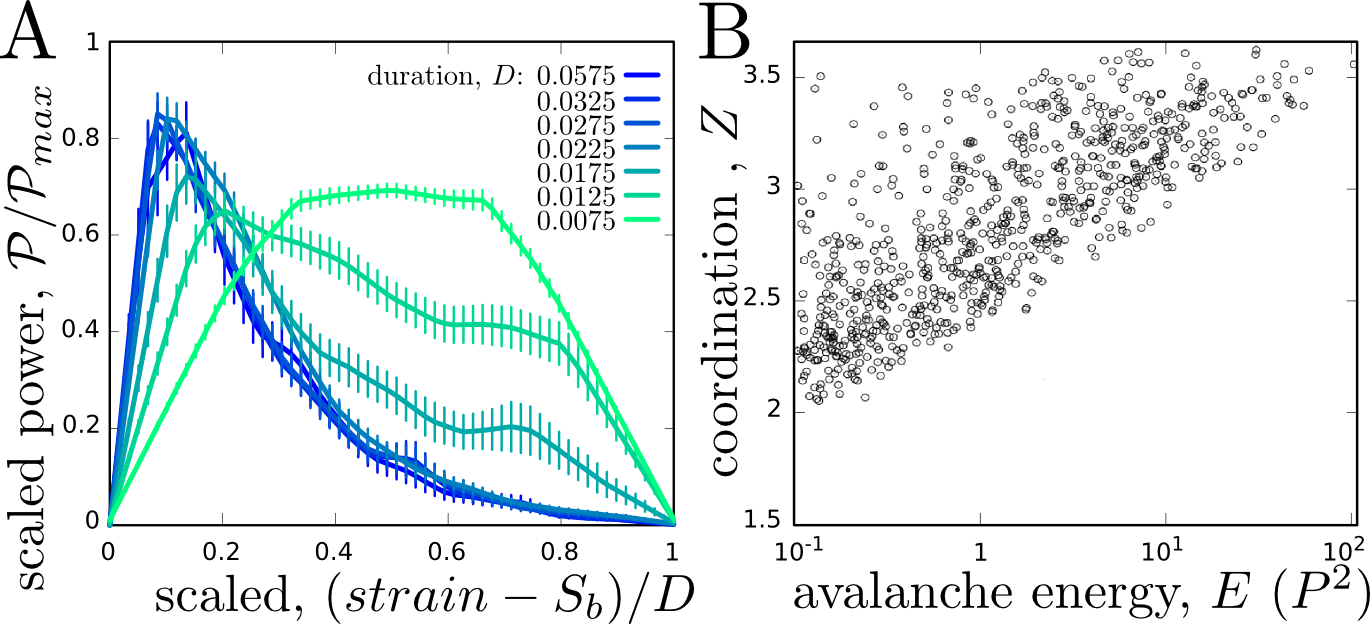}
\caption{(color online) A: Average `shape' of the global avalanches measured for different avalanche duration. Although avalanches are symmetric for short duration, they become progressively clearly asymmetric: a strong `acceleration' at the beginning and then a slow `deceleration'. B: Average coordination number $Z$ when an avalanche begins as a function of the energy $E$ of the avalanche. Avalanches happen equally in the shear and fragile regime, at both low and high coordination numbers. Both figures are plotted with data from experiment $I$.}
\label{figShapeZE}
\end{figure}

We also determine the average temporal avalanche shape for experiments $I$ of Table~\ref{tabExpe}. The avalanche shape provides a useful characterization of the avalanche/crackling dynamics and has been measured in a variety of systems \citep{zapperi_nat2005,Mehta_pre2006,laurson_pre2006,Papanikolaou_nat2011,danku_prl2013,bares_prl2015}. We adopt the standard procedure. First, we identify all avalanches $i$ of a given duration $D_i$; and second, we average the shape $\mathcal{P}(S \in [S_{b_i}, S_{e_i}]) / \max(\mathcal{P}(S \in [S_{b_i}, S_{e_i}])) ~ vs. ~ S/D_i$ over all avalanches, $i$. Fig.\ref{figShapeZE}-A shows the resulting shape and its dependence on duration $D$. For short avalanches (small $D$), the average shape is symmetric, but for longer avalanches (large $D$), it evolves toward a maximum energy loss near the beginning of the avalanche, followed by a slower rate of energy loss.

Fig.\ref{figShapeZE}-B shows the energy $E$ released during one avalanche $vs.$ the average coordination number $Z$ at the beginning of this avalanche. The correlation between both quantities is $41\%$ so, no correlation can be drawn between $E$ and $Z$ which means avalanches happen equally in the fragile ($Z<3$) and jammed regime ($Z>3$). Nevertheless, we observe an asymmetry between high energy-low coordination and low energy-high coordination:

\begin{equation} \label{eqRelaZE}
Z \geq 0.18 \cdot \log_{10}(E) + 1.9
\end{equation}

\noindent This relation implies that smaller avalanches occur at lower $Z$, corresponding to the fragile regime or the beginning of the jammed regime. Larger avalanches, including ones that are nearly system spanning, occur at or above the jammed region. This means that the coordination number has a particularly strong effect on the upper cut-off of the avalanche energy PDF, and a weaker effect on the lower cut-off.

\begin{figure}[htb!]
		\includegraphics[width=0.48\textwidth]{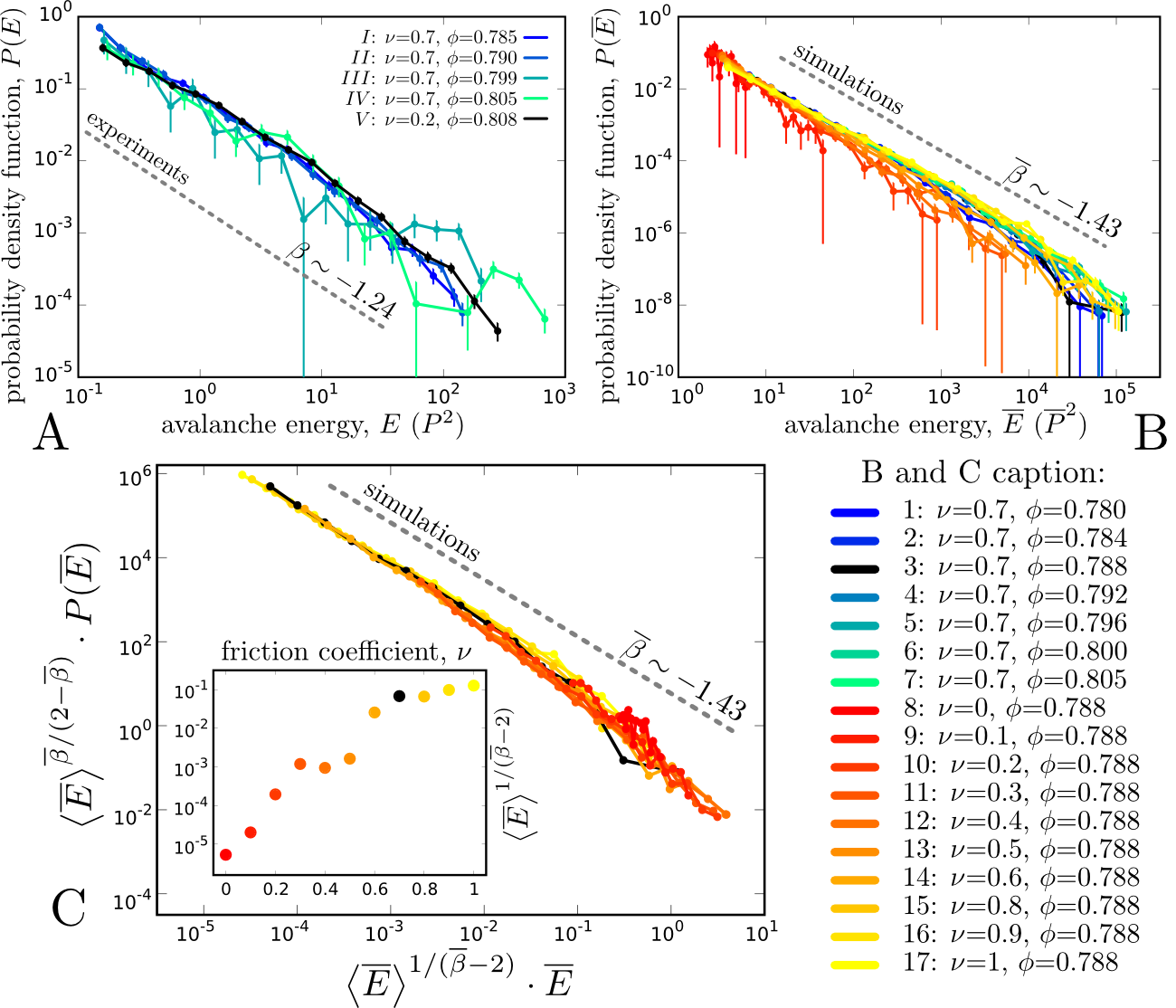}
\caption{(color online) A,B: Probability density function of the energy of global avalanches $P(E)$ measured in experiments and numerical simulations respectively. The particle packing fraction $\phi$ and the static friction coefficient between particles $\nu$ is varied over the different experiments and simulations. $P(E)$ follows a power-law with a constant exponent $\beta=-1.24 \pm 0.11$ for experimental data and $\overline{\beta}=-1.43 \pm 0.14$ for the numerical simulations. $\phi$ has no effect on the statistics while the upper cut-offs of the power-laws decrease with the friction coefficient $\nu$. In order to characterize the effect of the friction, $P(\overline{E})$ is rescaled using the method described in \citep{Patinet_prb2011} (C) to collapse curves for different friction and upper cut-off. C-inset: Scaling of the upper cut-off plotted as a function of $\nu$.}
\label{figPdfEVary}
\end{figure}

In the results below, we probe packing fractions spanning $0.785 \leq \phi \leq 0.808$, where throughout all experiments and simulations we keep the ratio of small to large particles constant. These packing fractions are below the frictionless isotropic jamming point $\phi < \phi_J$, but necessarily larger than the lower limit of shear jamming, $\phi > \phi_S$. We present $P(E)$ for each density in fig.\ref{figPdfEVary}-A. We note that in the experiment, the maximum shear amplitude is adjusted to avoid out-of-plane buckling of the particle layer. We do not see any significant effect of $\phi$ on $P(E)$, which means that if the maximum total stored energy $\mathcal{E}_{max}$ is higher for higher $\phi$, it does not change the value of the largest avalanches. As presented in fig.\ref{figPdfEVary}-B, we arrive at the same conclusion for the simulations, where we consider $\phi$'s spanning $0.780 \leq \phi \leq 0.805$; the packing fraction does not change the avalanche energy distribution.

We next consider the effect of varying the particle friction coefficient. We carried out two sets of experiments, one with friction coefficient $\nu=0.7$ and the other with $\nu=0.2$. As presented in fig.\ref{figPdfEVary}-A, there is no noticeable difference in $P(E)$ for the different $\nu$'s. In the numerical simulations, we varied $\nu$ over a broader range: from $\nu=0$ (no friction) to $\nu=1$. The overall scale of the PDFs, as measured by $P(\overline{E})$ vary with the friction coefficient $\nu$, while the overall shape and power-law exponent are insensitive to changes in $\nu$. In fig.\ref{figPdfEVary}-B, we show that the upper cut-off of the energy distribution power-law increases with $\nu$ for the simulations. This is explained by the fact that the Reynolds pressure increases more slowly with shear strain for low friction than for high friction, since grain scale particle rearrangements occur more easily for lower $\nu$. To quantify this effect, $P(\overline{E})$ in fig.\ref{figPdfEVary}-C is rescaled using the method presented in \citep{Patinet_prb2011}. For simulations with different friction coefficients, energy PDFs collapse on a single power-law with an exponent $\overline{\beta}=-1.43$ and an upper cut-off of a given shape. According to \citep{Patinet_prb2011}, the cut-off $\overline{E}_0$ scales with $\left\langle \overline{E} \right\rangle ^{1/(\overline{\beta}-2)}$. As presented in the inset of fig.\ref{figPdfEVary}-C, this quantity exhibits a sharp increase with increasing $\nu$, and the functional form is neither an exponential function nor a power-law.
 
\subsection{Local avalanches} \label{resLocal}

For a typical experiment, we investigate the effect of the different measures for avalanches at the local scale on the statistical behavior of the system. For each loading step, a grain is considered to be involved in an avalanche if one of the following criteria is satisfied:
\begin{itemize}
	\item Rotation from one step to another is larger than $8^\circ$;
	\item Non-affine displacement from one step to another is larger than $1.2{\rm~mm}$ ;
	\item Energy drop or rise is larger than $60$ ($G^4$);
	\item Pressure drop or rise is larger than $3$ ($G^2$).
\end{itemize}

\begin{figure}[htb!]
		\includegraphics[width=0.38\textwidth]{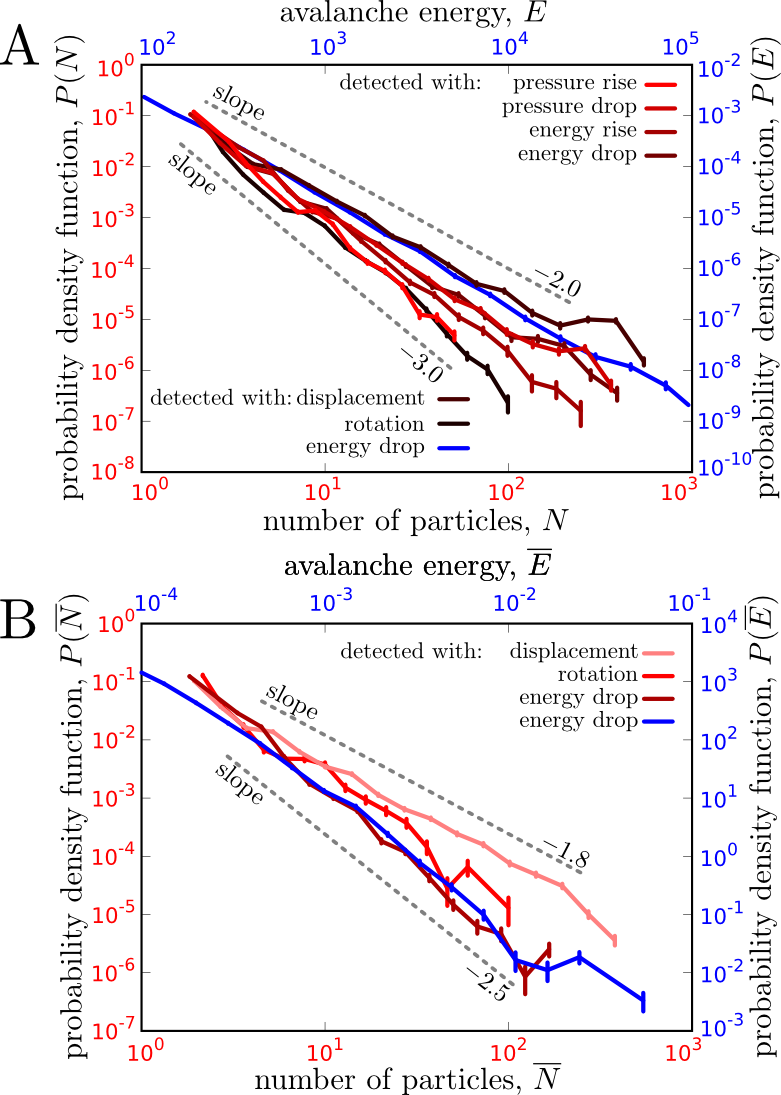}
\caption{(color online) A: Probability density functions $P(N)$ of the avalanche sizes measured at the local scale in terms of number of particles involved in each avalanche for experiment $I$. Avalanches are detected using a threshold on rotation, displacement and pressure/energy drop/increase. Power-laws are found with exponents varying from $-2$ to $-3$ (see text for details). B: Probability density functions $P(\overline{N})$ of the avalanche sizes measured in terms of number of particles involved in each avalanche for simulation $2$. Avalanches are detected using a threshold on rotation, displacement and energy drop. Power-laws are found with exponents varying from $-1.8$ to $-2.5$. In all cases the probability density function of the avalanche energies $P(E)$ detected from the energy drop is also plotted for comparison. A power-law spanned over almost $3$ decades is observed with exponent $\beta_l=-2.05 \pm 0.09$ for experiment and $\overline{\beta}_l=-2.08 \pm 0.1$ for numerical simulation.}
\label{figPdfENVaryLoc}
\end{figure}

\noindent These thresholds have the same effect on the statistics of the local avalanches as the energy threshold has on the statistics of the global avalanches, namely, it shifts the lower cutoff. Hence, we have chosen these threshold values so that the power-laws display the maximum number of decades. We then determined avalanche sizes in terms of $N$, the number of grains involved in the avalanches determined by the different physical quantities ({\it i.e.} angle, displacement, energy and pressure). The PDFs $P(N)$, presented in fig.\ref{figPdfENVaryLoc}-A, display power-laws over two to three decades, with exponents that depend on the physical quantity used to define the avalanche. Measurements of particle rotation or pressure rise lead to similar PDFs with exponents $-3.24 \pm 0.16$ and $-3.29 \pm 0.15$ respectively. PDFs based on pressure and energy drops are also power-laws, but with exponents $-2.30 \pm 0.13$ and $-2.26 \pm 0.10$ respectively. For avalanches based on energy increases, we find a power-law with exponent $-2.70 \pm 0.11$, which differs from all other exponents measured. Finally, the exponent based on the displacement ($-1.89 \pm 0.12$) is the same, within error bars, of the exponent based on energy drops ($\beta_l=-2.05 \pm 0.09$).

We also measured similar distributions in the simulations, but focused on avalanches defined by energy drops (threshold $4 \cdot 10^{-5}$ $P^2$), rotation (threshold $8^\circ$) and displacement (threshold $0.5$mm). Fig.\ref{figPdfENVaryLoc}-B shows that as in the experimental study, the power-law exponents depend on the quantities that are used to define them. Avalanches for displacements and rotations of particles display a power-law with exponents $1.82 \pm 0.10$ and $2.32 \pm 0.17$ respectively. For displacements, the simulations and experiments yield similar exponents. For avalanches determined from energy drops, the exponent is $-2.70 \pm 0.13$ if the avalanche size is given in terms of number of particles, while the exponent is $\overline{\beta}_l=-2.08 \pm 0.1$, similar to the value obtained in experiments. For the remainder of this paper, we will consider avalanches based on energy drops. In Table \ref{tabExpo} we summarize all the exponents measure from local avalanches.

\begin{table}[htb!]
	\begin{scriptsize}
		\begin{tabular}{|c|c|c|c|c|c|c|c|c|}
		\hline 
		Exponents & \multicolumn{2}{c|}{experiments} & \multicolumn{2}{c|}{simulations}  \\ 
		\hline 
	\begin{tabular}{c} measured\\from field: \end{tabular} & \begin{tabular}{c} measured by\\field intensity\\variation \end{tabular} & \begin{tabular}{c} measured\\by number\\of particles \end{tabular} & \begin{tabular}{c} measured by\\field intensity\\variation \end{tabular} & \begin{tabular}{c} measured\\by number\\of particles \end{tabular} \\ 
		\hline 
		displacement & $-1.80 \pm 0.05$ & $-1.89 \pm 0.12$ & $-1.63 \pm 1.2$ & $-1.82 \pm 0.10$ \\ 
		\hline 
		rotation & $-2.21 \pm 0.11$ & $-3.24 \pm 0.16$ & $-1.62 \pm 0.09$ & $-2.32 \pm 0.17$ \\ 
		\hline 
		energy drop & $-2.05 \pm 0.09$ & $-2.26 \pm 0.10$ & $-2.08 \pm 0.1$ & $-2.70 \pm 0.13$  \\ 
		\hline
		energy rise & $-2.35 \pm 0.14$ & $-2.70 \pm 0.11$  \\ 
		\cline{1-3} 
		pressure drop & $-2.22 \pm 0.15$ & $-2.30 \pm 0.13$  \\ 
		\cline{1-3}
		pressure rise & $-2.15 \pm 0.17$ & $-3.29 \pm 0.15$  \\ 
		\cline{1-3}
		\end{tabular} 
	\end{scriptsize}
\caption{Exponents of the probability density functions of avalanche sizes measured at the local scale in terms of number of particles involved in avalanche and total variation of the field considered to detect the avalanche, for experiment $I$ and numerical simulations $2$.}
\label{tabExpo}
\end{table}

\begin{figure}[htb!]
		\includegraphics[width=0.31\textwidth]{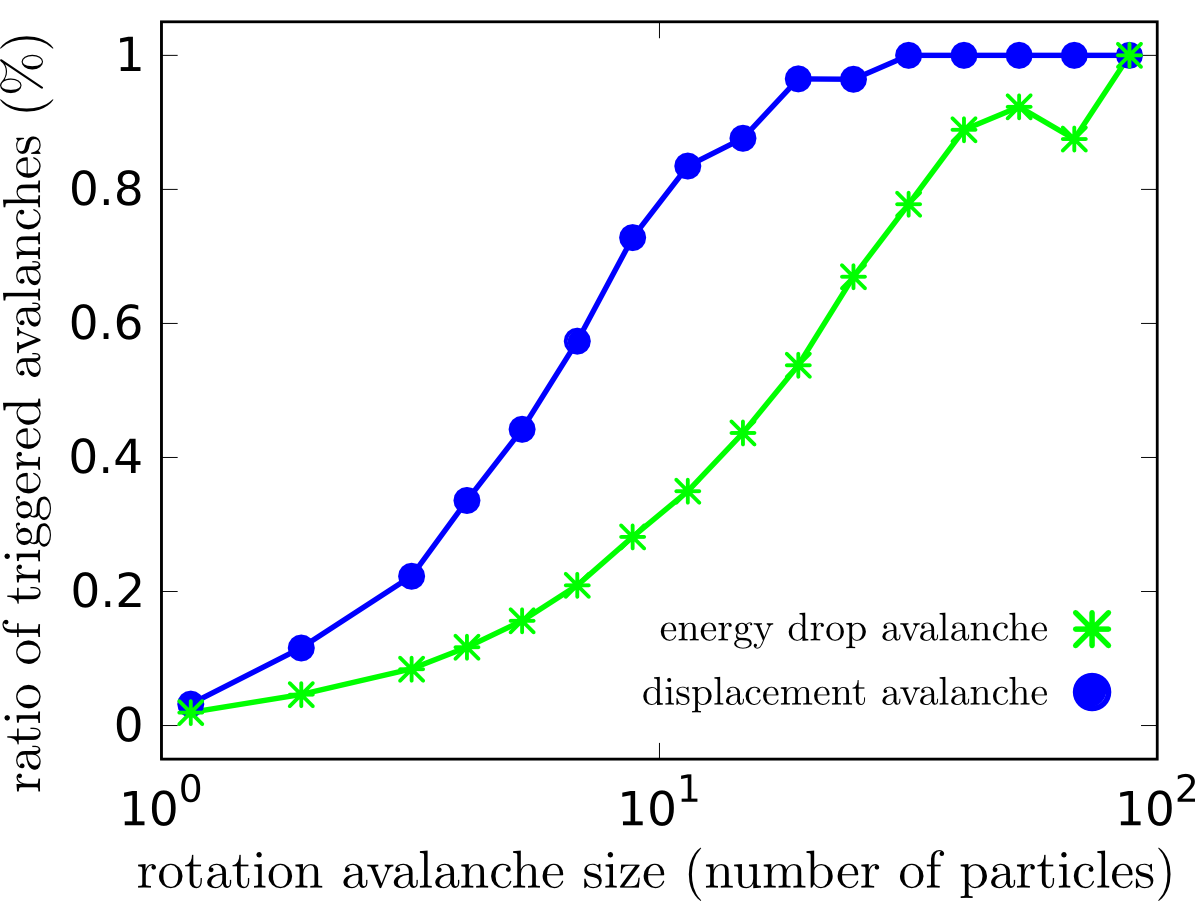}
\caption{(color online) For experiment I, we plot the ratio of the number of avalanches based on the rotation to the number of avalanches based on the displacement (or energy drop) provided the avalanches have an overlap in space and time. (more details in the text).}
\label{figratioTrigger}
\end{figure}

As shown in fig.\ref{figLocalAngleDisp} and \ref{figLocalExtraction}, avalanches based on local particle rotation, local displacement, or local pressure drop tend to be clustered in space and time. Thus, avalanches in local particle rotation, displacement, and pressure drop represent similar types of rearrangements. However, this tends not to be the case for small rearrangements where few particles move or rotate. Fig.\ref{figratioTrigger} shows data for all avalanches based on particle rotation involving a given number of grains. For these avalanches, we determine the ratio of avalanches that are also associated with an avalanche based on the displacement or energy drop. `Associated' means that among the grains involved in the rotation avalanche, there is at least one which is also involved in a simultaneously occurring displacement or energy drop avalanche. Fig.\ref{figratioTrigger} shows that most of the small avalanches in rotation are independent of any other kind of rearrangement. However, the biggest avalanches involve rearrangement in rotation, displacement and stored energy for the particles involved. Avalanches involving more than $20$ grains are always detected based on both rotation and displacement, but not necessarily on the variation of the energy. This is mostly due to the fact that rearrangements of particles involving translation and rotation need not involve strong contacts with other grains.

\begin{figure}[htb!]
		\includegraphics[width=0.5\textwidth]{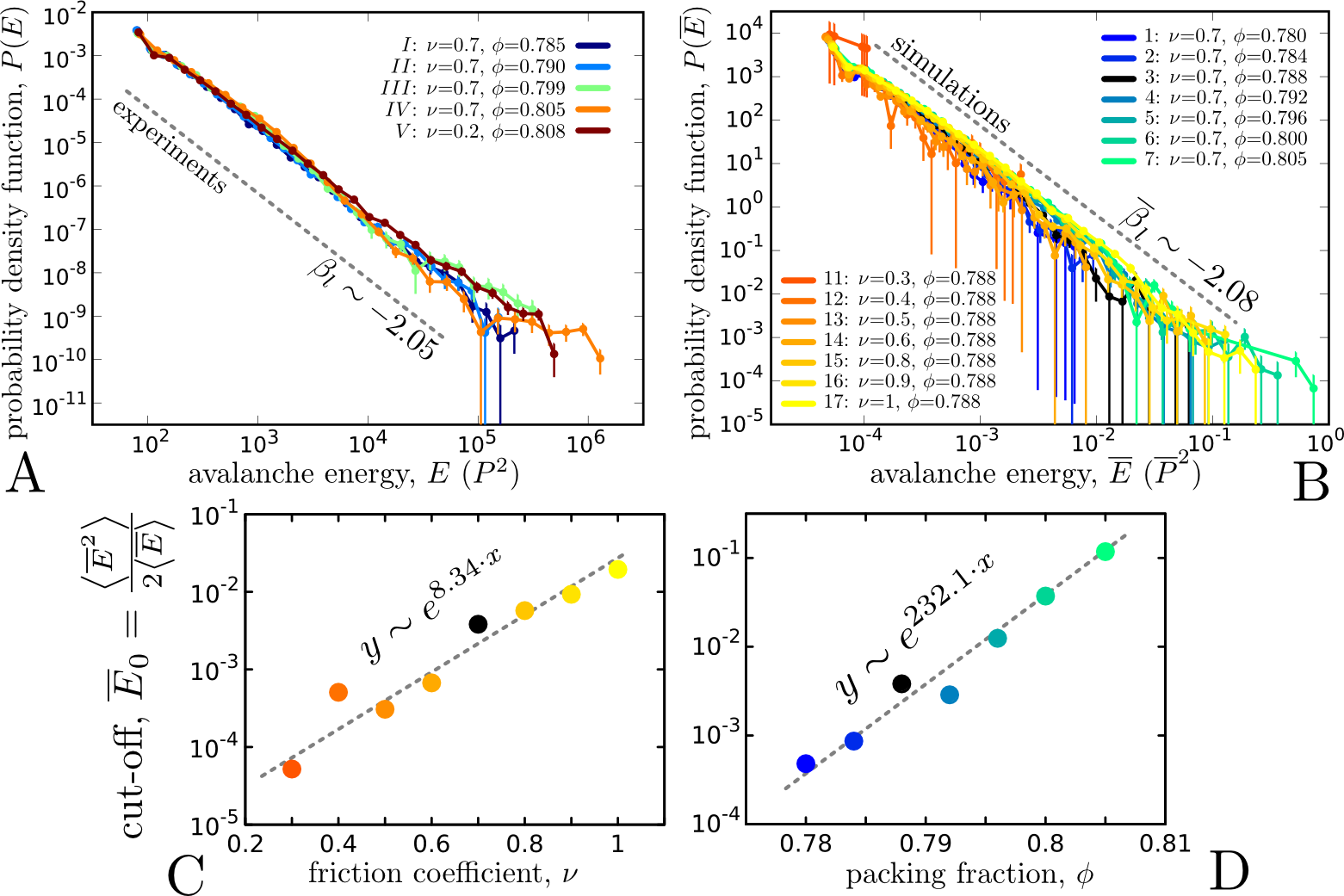}
\caption{(color online): Probability density function of the avalanche energy $P(E)$ detected from the energy drop for different packing fractions $\phi$ and particle friction coefficients $\nu$ in experiments (A) and simulations (B). Within the error bars, power-laws with similar exponents  are observed. In C and D, we show the variation of the power-law upper cutoff ${\bar E}_0$ as a function of the inter-particle friction $\nu$ at $\phi = 0.788$ and as a function of the packing fraction $\phi$ at $\nu = 0.7$.}
\label{figPdfEVaryLoc}
\end{figure}

As for global avalanches, we now investigate the effects of the packing fraction, $\phi$, and of the inter-particle friction coefficient, $\nu$, on local avalanche energy PDFs. Fig.\ref{figPdfEVaryLoc}-A shows experimental results for different $\phi$ and for particles that were/were-not wrapped with Teflon$^{\mbox{\scriptsize{\textregistered}}}$ tape. As for global avalanches, the PDFs follow power-laws and the exponents are unaffected by the variations in the packing fraction and friction coefficient. Only the upper cut-offs change with $\phi$ or $\nu$. Simulations, shown in fig.\ref{figPdfEVaryLoc}-B, have similar power-law exponents as the experiments, but the upper cut-off $\overline{E}_0$ depends on $\phi$ and $\nu$. Fig.\ref{figPdfEVaryLoc}-C and D shows the upper cut-off determined using the scaling explained in \citep{rosso_prb2009}. This upper cut-off increases with both $\nu$ and $\phi$ and obeys an exponential form: 

\begin{equation} \label{eqCutoffPdfEVarNu}
	\overline{E}_0 = \dfrac{\left\langle \overline{E}^2 \right\rangle}{2 \left\langle \overline{E} \right\rangle} \sim e^{(8.34\pm1.2) \cdot \nu + (232.1\pm17) \cdot \phi}
\end{equation}

\begin{figure}[htb!]
		\includegraphics[width=0.5\textwidth]{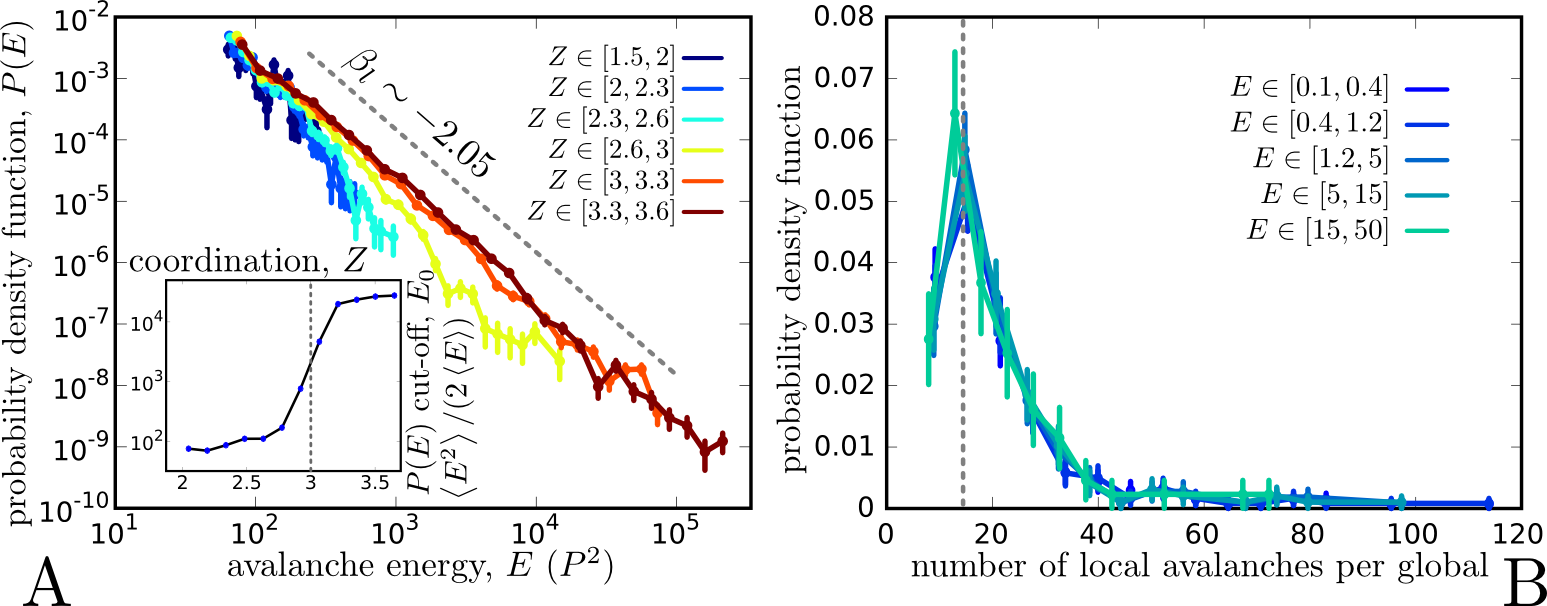}
\caption{(color online) A: For experiment $I$, probability density function of the avalanche energy, $P(E)$, detected from the energy drop for different coordination numbers $Z$. A-inset: Evolution of the energy upper cut-off $E_0$ as a function of the coordination number $Z$. A sharp increase of $E_0$ is observed near $Z=3$, which is the minimal number of contacts for large-scale mechanical stability. B: For experiment $I$, probability density function of the number of local avalanches detected during a global avalanche for different ranges of global avalanche energy, $E$.}
\label{figPdfGlobalLocalPdfEZ}
\end{figure}

We now consider local avalanche statistics as a function of Z. For each avalanche of size $E_i$, a coordination value $Z_i$ is attributed by measuring the average contact number per grain for the whole packing at the beginning strain $S_{b_i}$ of the avalanche: $Z_i=Z(S_{b_i})$. Fig~\ref{figPdfGlobalLocalPdfEZ}-A, which gives energy PDFs for different $Z$ shows that the range of $Z$  has no effect on the power-law exponent for the distribution of energy drops. Nevertheless, it clearly has an effect on the upper cut-off $E_0$, which increases with the coordination number, as shown in the inset of Fig.\ref{figPdfGlobalLocalPdfEZ}-A. Higher particle coordination corresponds to higher pressure and larger variations in energy and other quantities. We also note that higher $E_0$ corresponds to values of $Z$ such that $Z>d+1$ with $d=2$ the system dimension. This means that when the system is jammed, the upper energy cut-off is distinctly larger.

To quantify the link between local and global avalanches, and to understand how local avalanches induce global ones, we show in fig.\ref{figPdfGlobalLocalPdfEZ}-B the PDF of the number of local avalanches included in a global avalanche. For each global avalanche of experiment $I$ whose energy $E$ is in a given range, we count the number of local avalanches between the beginning and end of the global avalanche. The PDF of this number of local avalanches contained in a global avalanche is shown in Fig.\ref{figPdfGlobalLocalPdfEZ}-B over a range of avalanche energies. The average value of local avalanches per global avalanche, $\sim 15$, is surprisingly independent of the energy range of the global avalanche. Hence, whatever the size of a global avalanche, on average, it is composed of the same number of local avalanches.

\begin{figure}[htb!]
		\includegraphics[width=0.4\textwidth]{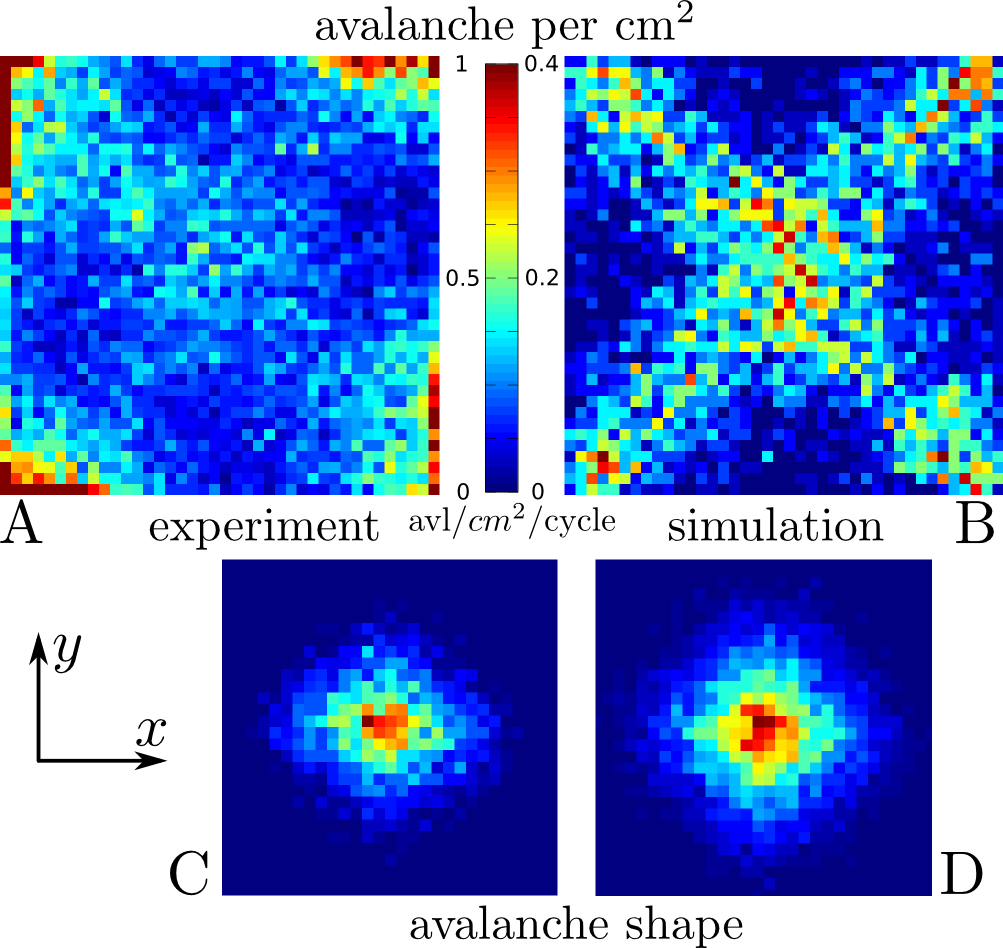}
\caption{(color online) A, B: Number of local avalanches detected with particle displacement per unit area per cycle for experiment $I$ and simulation $2$ respectively. C, D: Average local avalanche shape, for avalanches detected with the energy drop with size $N$ between $8$ and $10$ particles for experiment $I$ and simulation $2$ respectively. See text for more details.}
\label{figPdfDensiAvlShape}
\end{figure}

Finally, we consider the location and shape of the local avalanches. To determine whether avalanches occur homogeneously throughout the whole system, we compute their non-affine position, defined as their actual position from which is substracted the associated affine displacement, and determine the number of avalanches per unit area per shear cycle. The map of this quantity is shown in fig.\ref{figPdfDensiAvlShape}-A and B for experiment $I$ and simulation $2$ respectively. For better accuracy of the avalanche position, we consider only avalanches detected by thresholding the local displacement. In both experiments and simulations, the highest density occurs along diagonals which correspond to shear bands. This means that avalanches are mainly triggered in this area which corresponds to highly turbulent granular flow. Also, in the experimental case, we observe an abundance of avalanches triggered near the edges of the experimental cell. We believe this is due to the small loading imperfections caused by the gap between boundary walls and basal plates and between crossing boundary walls.

Fig.\ref{figPdfDensiAvlShape}-C and D shows the average 3D shape of the avalanches computed for avalanches detected by energy drop in experiment $I$ and simulation $2$. For each avalanche involving $8$ to $10$ grains which corresponds to the average avalanche size), the shape of the avalanche matches the local density variations of the grains. This is obtained by considering only particles involved in a given avalanche, choosing an origin at the barycentre of the avalanches ($x$-$y$ directions are kept the same as the ones given in fig.\ref{Setup}-B) and by computing the local density field. This field, which we refer to as `shape', is then averaged over several avalanches, where the origin for each shape is its center of mass. Fig.\ref{figPdfDensiAvlShape}-C and D show the result in both experiments and simulations. The average shape is not isotropic, and avalanches are preferentially aligned along the instantaneous compressive shear direction, which in turn corresponds to the strong force chain direction. 

\section{Discussion and Conclusions} \label{discussion}

We have observed scale free distributions for global avalanches based on the power dissipated in both experiments and related simulations. The power-law exponents from these two different approaches agree within statistical errors. We find an exponent associated with energy release per avalanche of $\beta = - 1.24 \pm 0.11$ for experiments and $\beta = -1.43 \pm 0.14$ for simulations. These are in agreement with the $1.36$/$1.34$ theoretical results reported by \citep{lin_pnas2014} for sheared soft particles. However they are somewhat smaller than the $1.5$ exponent computed for \citep{dahmen_nat2011}'s discrete model or the $\sim 1.5$ exponent measured experimentally by \citep{kabla2007_jfm} for sheared foam and by \citep{Hayman2011_pag} for 2D granular material. We note that the simulations of \citep{dahmen_nat2011} carried out with an elasto-plastic cellular automaton model including an Eshelby stress redistribution kernel, do not include some of the distinguishing physical processes present in our experiment. Perhaps most importantly, static friction and force chains are not part of this model but are key features of our system. The studies in \citep{kabla2007_jfm} consider highly deformable particles with a very low friction coefficient, which may be a key difference from our system. Finally, the exponent reported in \citep{Hayman2011_pag} has a rather large errorbar, and it may be consistent with our results, even if the loading mechanism is different. In contrast to \citep{dahmen_nat2011}, the discrete dislocation model of \citep{Ispanovity2014_prl} predicts an exponent of $1$. However, this model also does not include effects associated with friction or force chains, and the system is stress controlled.

We find a power-law distribution for the depinning energy that differs from what is observed in pinning-depinning models. A possible cause for this difference may lie in the fact that in the granular case, the quenched disorder in the system changes after each avalanche. Other possible causes may be the inherent anisotropy and qualitatively different elastic response preceding an avalanche. Hence, the granular case may differ at a very basic level from the ingredients of models that have been constructed for other types of amorphous materials \citep{dahmen_nat2011,lin_pnas2014,weiss2014_pnas,regev2015_nat,Tyukodi2015_arx}. As a consequence, the plasticity for dense granular systems appear to differ qualitatively and quantitatively from the behavior of pinning-depinning model \citep{Ispanovity2014_prl}. 

We find that the temporal avalanche shape can differ significantly from numerical models \citep{dahmen_nat2011,aharonov_jgr2004}, with a clear asymmetry of the shape for long lasting avalanches. We believe this is mainly due to the fact that, in our system, the basal static friction -- between the particles and the supporting glass -- induces a large viscosity in the particle displacement and reduces the intensity of their displacement at long terms. Nevertheless, this does not completely explain why the effect would be different for weak and strong avalanches and remains a hypothesis. Another hypothesis is can also explain asymmetry for large avalanches. In large events, the system unloads from a state of high over-compression to one of low compression. In small events, the amount of unloading may be smaller. For small unloading, the system is locally harmonic, and the slip event will be symmetric. However, for large events, the system is initially very stiff, and then softens as unloading occurs. We would expect that this would lead to rapid dynamics initially \textit{i.e.} the effective elastic coefficient is large, hence the early peak in the response. As the system continues to unload, The system softens and the time scale for relaxation grows towards the end of the avalanche. This would lead to a weak extended feature at the end of the avalanche. From this perspective, the asymmetry for large events arises in the nonlinear elasticity of the system. A last possible origin of the avalanche shape anisotropy is the strong correlation between avalanche dynamics and force networks. 

We find that the statistical behavior of the local and global avalanches is independent of the jamming regime as long as force chains are present. Only non-universal parameters, such as power-law cut-offs, depend on the granular regime (\textit{e.g.} coordination number and packing fraction). The upper cut-off changes strongly and nonlinearly when the jamming transition is crossed; \textit{i.e.} avalanches are larger when the system is jammed. Similarly, the packing fraction and the inter-particle friction coefficient have no effect on the power-law exponents of the avalanche PDFs but do have an effect on their upper cut-off. For local avalanches, the upper cut-off increases exponentially with the friction coefficient and the packing fraction, while at the global scale it increases only with $\nu$ but not with $\phi$ contrary to what is predicted by the class of models considered by Dahmen et al. \citep{dahmen_nat2011}. We also find a difference between numerical and experimental data on this point: the upper cut-off changes when friction and packing fraction change in numerical simulation but not in experiments for both the local and global scales. Since the main difference between the numerical model and the experimental set-up is the basal friction which is mimicked numerically by adding a strong viscous damping, we believe the upper cut-off is dominated in the experimental case by the basal friction phenomenon that is why it does not vary.

We introduce a novel approach for avalanche data analysis by extracting avalanches from the local scale of several different observables: displacement, rotation, pressure and energy fields. We find different statistical behavior, \textit{i.e.} power-law exponents, for avalanches detected using these different physical quantities. Specifically, distributions for displacement and rotation are not the same, and distributions for pressure/energy differ for loading and unloading.  Results from our experiments and simulations differ from a number of models \citep{regev2015_nat,maloney2006_pre} and from results obtained by shearing a soft slippery granular medium \citep{kabla2007_jfm}, where stress is redistributed according to a symmetric Eshelby kernel, and displacements follow a T1-type rearrangement dynamics. We believe that the key difference between these models and the present granular experiments, and presumably other dry granular systems, is that the granular systems form strong anisotropic force networks in response to shear. These structures are key features of granular systems, and they are not part of typical models. To our knowledge, they do not occur for soft slippery materials. The fact that the spatial structure of granular avalanches tends to be elongated along the compression direction is one obvious indicator of how force chain structures impact granular dynamics, and consequently, the avalanche statistics.

Finally, we find that on average, the number of local per global avalanches is constant whatever the energy of the global avalanche. This result suggests that the local structure of a global avalanche is statistically independent of the global avalanche intensity. Also, as pointed out by \citep{kabla2007_jfm} we find that the triggering of avalanches is strongly coupled to the non-affine displacement of the particles in the shear band. Although the event may be triggered in the relatively weak shear band region, the fact that the system is everywhere close to force balance leading up to an avalanche means that the resulting stress drop can span the whole system.

\section{Acknowledgements} \label{acknowledgement}

We acknowledge support from the W. M. Keck Foundation Grant No. DTO61314 (T.B., J.B., R.B., and C.S.O.), National Science Foundation Grant Nos. DMR1206351 (J.B. and R.B.), DMS-1248071 (J.B. and R.B.) CMMI-1462439 (C.S.O.), NASA grant NNX15AD38G (J.B. and R.B.), and Labex NumEv anr-10-labx-20 (J.B.).

\appendix

\section{Definition of the elastic and viscoelastic coefficients of the contact force model}\label{app:coefficients}

In this section, we detail the relation between the elastic and viscoelastic constants involved in the model for the normal and tangential contact forces ($k_n$, $k_t$, $\gamma_n$, $\gamma_t$) and the grains material properties for particle $i$ (Young's modulus $E_i$, Poisson's ratio $\mu_i$, and coefficient of restitution $c_r$) used in the simulations. We consider a pair of interacting particles ($i$,$j$) with normal displacement vector $\delta \mathbf{n}_{ij}$. The elastic ($k_n$, $k_t$) and viscoelastic ($\gamma_n$, $\gamma_t$) coefficients from this pair of interacting particles are calculated as follows from the material properties:
\begin{align}
&k_n = \frac{4}{3} Y^* \sqrt{R^*\delta_n} \\
&\gamma_n = -2 \sqrt{\frac{5}{6}} \beta \sqrt{S_n m^*} \\
&k_t = 8 G^* \sqrt{R^* \delta_n}\\
&\gamma_t = -2 \sqrt{\frac{5}{6}} \beta \sqrt{S_t m^*}
\end{align}

\noindent where we denote: 
\begin{align}
&\delta_n = \left|\delta \mathbf{n}_{ij}\right| \\
&S_n = 2Y^*\sqrt{R^*\delta_n} \\
&S_t = 8G^*\sqrt{R^*\delta_n} \\
&\beta = \frac{\ln c_r}{\sqrt{\ln^2c_r + \pi^2}}
\end{align}

\noindent and where we have defined the mean Young's modulus $Y^*$, shear modulus $G^*$, radius $R^*$ and mass $m^*$ of the interacting pair by:

\begin{align}
& \frac{1}{Y^*} = \frac{1-\mu_i^2}{Y_i}+\frac{1-\mu_j^2}{Y_j} \\
& \frac{1}{G^*} = \frac{2(2+\mu_i)(1-\mu_i)}{Y_i}+\frac{2(2+\mu_j)(1-\mu_j)}{Y_j} \\
& \frac{1}{R^*} = \frac{1}{R_i}+\frac{1}{R_j} \\
& \frac{1}{m^*} = \frac{1}{m_i}+\frac{1}{m_j}
\end{align}


\bibliography{biblio} \label{biblio}

\end{document}